\documentclass[a4paper]{ar}
\usepackage[numbers]{natbib}
\usepackage{textcomp}

\jname{Xxxx. Xxx. Xxx. Xxx.}
\jvol{AA}
\jyear{YYYY}
\doi{10.1146/((please add article doi))}

\begin{document}

\markboth{Reichhardt et al.}{Ratchet effects in active matter systems}

\title{Ratchet Effects in Active Matter Systems}

\author{C.J. Olson Reichhardt and C. Reichhardt
  \affil{Theoretical Division, Los Alamos National Laboratory, Los Alamos, NM, USA, 87545;
    email: cjrx@lanl.gov}}

\begin{abstract}
  Ratchet effects can arise for single or collectively interacting Brownian particles on an asymmetric substrate when a net dc transport
  is produced by an externally applied ac driving force or by periodically flashing the substrate.  Recently, a new class of active ratchet systems has been realized through the use of active matter, which are self-propelled units that can be biological or non-biological in nature.  When active materials such as swimming bacteria interact with an asymmetric substrate, a net dc directed motion can arise even without external driving, opening a wealth of possibilities such as sorting, cargo transport, or micromachine construction.  We review the current status of active matter ratchets for swimming bacteria, cells, active colloids, and swarming models, focusing on the role of particle-substrate interactions.  We describe ratchet reversals produced by collective effects and the use of active ratchets to transport passive particles.  We discuss future directions including deformable substrates or particles, the role of different swimming modes, varied particle-particle interactions, and non-dissipative effects. 
\end{abstract}

\maketitle

\tableofcontents

\section{INTRODUCTION}

When a particle is placed in a symmetric periodic potential and subjected to an ac drive,
in general the particle simply rocks back and
forth and undergoes no net dc motion.
If an external dc drive $F^{dc}$ is imposed on a particle in a substrate, 
the particle has no net dc motion and remains pinned
as long as $F^{dc} < F_{c}$,
where $F_{c}$ is the critical depinning force for motion over the substrate.
Once $F^{dc} > F_{c}$, the particle enters a sliding state \cite{1}.  
When the motion of the particle is 
overdamped, its dynamics can be described by the following equation of motion: 
\begin{equation}
\eta \frac{d{\bf R}}{dt} = {\bf F}^{s} + {\bf F}_{n} + {\bf F}_{\rm ext} + {\bf F}^{T} 
\end{equation}
where $\eta$ is the damping constant,
${\bf F}^{s}$ is the force from the substrate,
${\bf F}_{n}$ is the interaction force from the other particles,
${\bf F}_{\rm ext}$ is an externally applied driving force,
and ${\bf F}^{T}$ is a stochastic force representing thermal fluctuations.

\begin{figure}[h]
\includegraphics[width=4in]{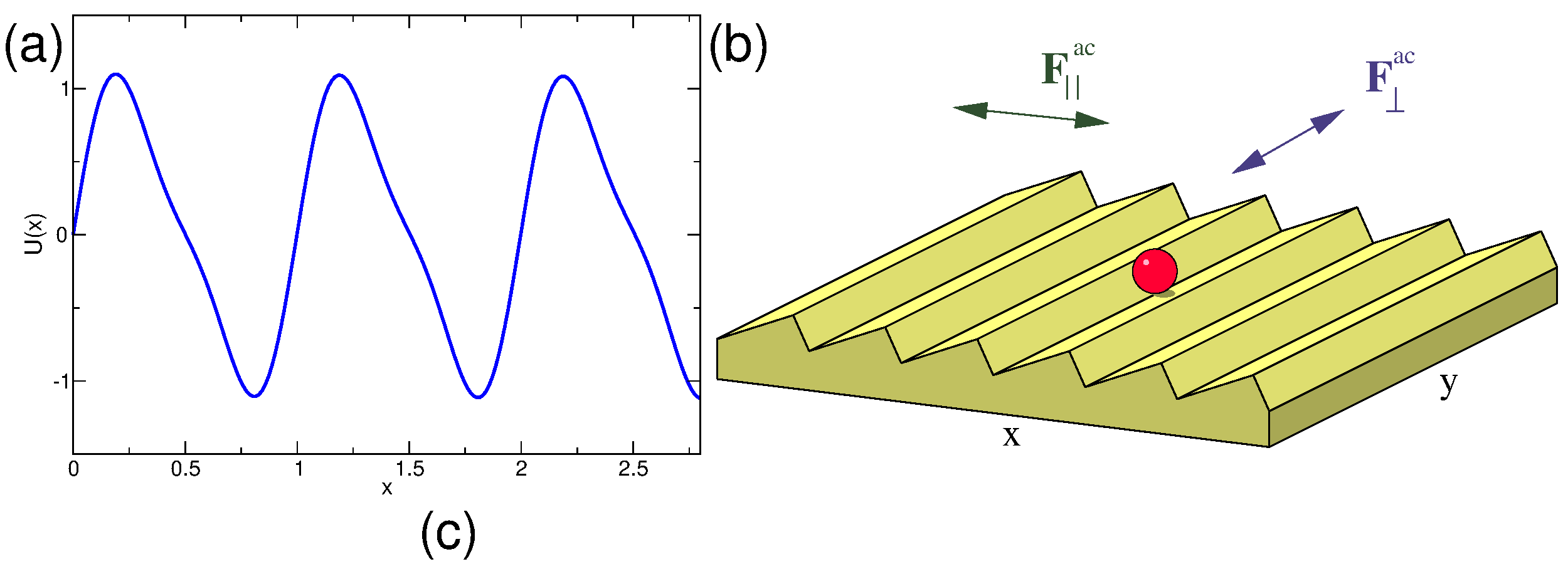}
\includegraphics[width=2in]{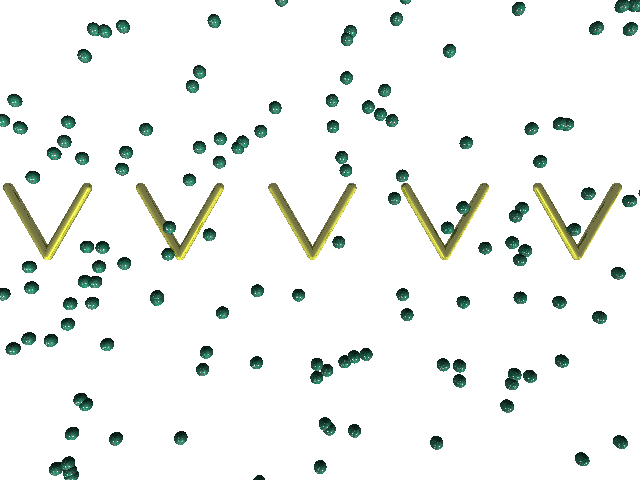}
\caption{ (a) A plot of an asymmetric one-dimensional substrate potential.
  Under an ac drive, the ratcheting motion of particles
  is in the easy direction of the substrate, corresponding to the negative
$x$ direction.
  (b) A quasi-1D version of the same substrate for particles moving in two
  dimensions.  Adapted from C. Reichhardt, D. Ray, and C.J. Olson Reichhardt,
  New J. Phys. 17, 070304 (2015).
  (c) A line of 2D funnels or barriers (yellow) affecting the motion of interacting
  particles (green).  The easy direction for the ratchet effect is the positive $y$ direction,
  toward the top of the panel.
}
\label{fig:1}
\end{figure}

The situation changes if the substrate is asymmetric.
An example of a one-dimensional (1D)
asymmetric substrate potential with periodicity $a$,
\begin{equation}
U(x) = U_{0}[\sin(2\pi x/a) + 0.25\sin(4\pi x/a)] ,
\end{equation}
is illustrated in Fig.~\ref{fig:1}(a).
In the absence of thermal fluctuations,
the maximum pinning force exerted by the substrate on
the particle for motion in the  
positive $x$-direction 
is $3\pi U_{0}/a$, while
the maximum pinning force
for motion in the negative $x$-direction
has a smaller value of $3\pi U_{0}/2a$. As a result,
if an externally applied ac driving force of the form 
${\bf F}_{\rm ext} = F_{AC}\cos(\omega t){\hat {\bf x}}$ is introduced,
then the particle remains trapped and oscillates within a single potential
minimum for
$F_{AC} < 3\pi U_{0}/2a$.
In contrast, when
$3\pi U_0/a \leq F_{AC} < 3\pi U_{0}/2a$, the particle can overcome
the substrate potential barrier for motion
in the negative $x$-direction but cannot overcome the barrier for motion in
the positive $x$-direction, so that during each ac cycle the particle
undergoes a net
dc transport in the negative $x$-direction.
Such motion is termed a ratchet effect.
The amount of dc translation
that occurs per cycle depends on the amplitude and frequency of the ac drive.
For fixed ac frequency, the efficiency of the ratchet
is generally nonmonotonic, increasing
with increasing $F_{AC}$ for $F_{AC}<3\pi U_0/2a$ and then decreasing with
increasing $F_{AC}$ at higher $F_{AC}$
\cite{2}.

When the mechanism responsible for producing a ratchet effect is
an ac driving force, the system is known as a rocking ratchet.
An alternative mechanism operates in a flashing ratchet, where
thermally diffusing particles are placed on a substrate
that is periodically flashed on and off, and where a maximum ratchet efficiency
emerges as a function of flashing frequency \cite{2}.
In the single particle limit for overdamped particles, the ratchet effect 
causes the particle to translate along the easy direction of the substrate regardless of
the nature of the ratchet.
When collective particle-particle interactions are introduced, however,
it is possible to observe
reversals of the ratchet effect,
where the motion is along the hard direction of the substrate asymmetry
\cite{2,3,4,5}.
These reversals are often produced by commensurability effects between
the interacting particles and the substrate, where emergent quasiparticle objects
such as kinks or antikinks dominate the transport and behave as if they are moving
through an inverted substrate potential.
Examples of higher-dimensional asymmetric substrates on which ratchet effects can occur
appear in Fig.~\ref{fig:1}(b), which shows a quasi-1D asymmetric potential with no structure
transverse to the asymmetry direction \cite{5,6},
and in Fig.~\ref{fig:1}(c), which shows a linear
array of V-shaped barriers \cite{7}.
The asymmetry necessary for generating ratchet effects
can also be introduced by placing individual symmetric pinning sites in
an array that has a density gradient along one direction \cite{8}.
In geometric or drift ratchets, which require a minimum of two spatial dimensions,
particles driven over an asymmetric substrate by
a dc drive undergo a net dc drift in the direction transverse to the applied dc drive
\cite{9,10,11}.
In transverse ratchets, which are a variation of drift ratchets, a net dc ratchet motion
arises in the direction transverse to an applied ac driving force \cite{12,12N,14}.
Additional ratchet effects including ratchet reversals can arise
when there are non-dissipative terms in the equations of particle motion,
such as inertia \cite{15} or a Magnus term \cite{16}.   

Studies of ratchet effects have been performed
for colloids interacting
with asymmetric flashing substrates \cite{17},
vortices 
in type-II superconductors interacting with
asymmetric periodic 1D \cite{5,6} and 2D substrates
\cite{4,12,12N,14}, 
granular matter on sawtooth substrates \cite{18,19},
cold atoms \cite{20,21},
electron systems \cite{22,23}, 
and domain walls moving over asymmetric substrates \cite{24,25}.
These systems all require application of some form of external driving
in order to create the nonequilibrium conditions
necessary for a ratchet effect to occur.
In a different
class of nonequilibrium systems, the dynamics is governed by
internal driving or self-propulsion \cite{26,27,28,29}. 
These active matter systems include
flocking and swarming particles, pedestrian and traffic flow, swimming bacteria,
crawling cells, and actin filaments.
Active matter has been 
attracting growing attention due to the increasing availability of
non-biological active systems such as artificial swimmers, self-driven colloids,
robot swarms, and autonomous agents.
Such systems exhibit a variety of phenomena 
including pattern formation \cite{30}, 
self-clustering \cite{31,32,33,34},
and nonequilibrium phase transitions \cite{35},
and represent a rich and very active field of research.

\section{RATCHET EFFECTS FOR SWIMMERS IN FUNNEL ARRAYS}
\begin{figure}[h]
\includegraphics[width=3in]{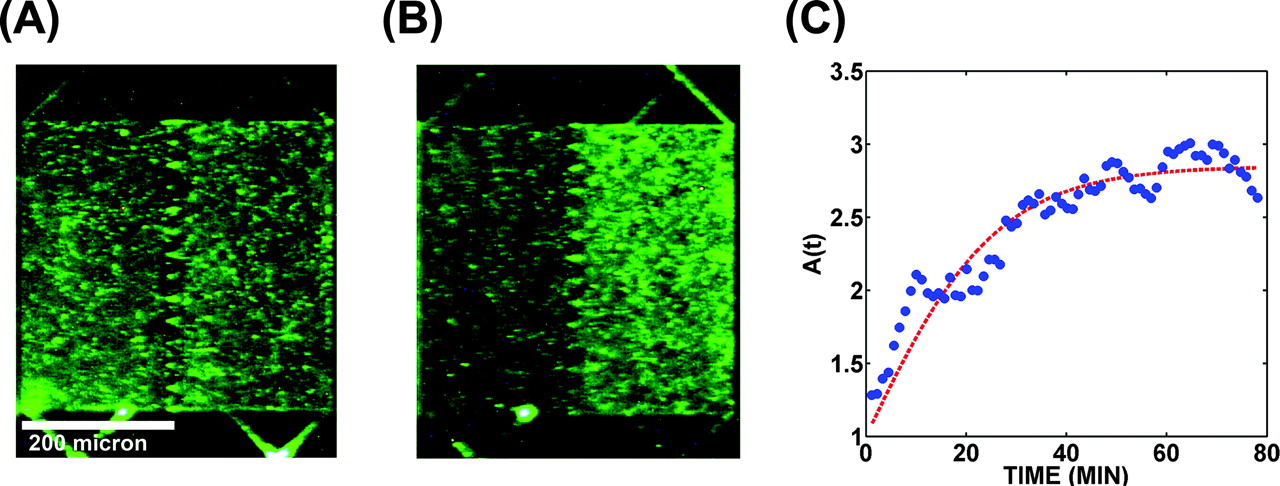}
\caption{(a,b)
  Experimental images of the concentration of bacteria, indicated by fluorescence,
  in two chambers separated by a series of
  funnel barriers resembling those illustrated
  in Fig.~\ref{fig:1}(c). The easy-flow direction through the barriers is toward the
  right chamber.  (a) The initial uniform distribution of bacteria at time $t=0$. (b) 
  The distribution at $t=80$ minutes showing that the bacteria concentration is
  significantly higher in the right chamber.
  (c) The ratio $A=\rho_R/\rho_L$ 
  of the bacteria density $\rho$ in the right and left chambers
  versus time, showing that a rectification effect is occurring.
  Reprinted with permission from
  P. Galajda, J. Keymer, P. Chaikin, and R. Austin,
  J. Bacteriol. 189, 8704 (2007).  Copyright 2007 by the American Society for Microbiology.
}
\label{fig:2}
\end{figure}

One of the first demonstrations of an active matter ratchet was
achieved by P. Galajda {\it et al.} \cite{7} for a system of {\it E. coli} swimming
in a sample divided into two chambers by an array of funnel-shaped barriers
similar to those illustrated in Fig.~\ref{fig:1}(c).
The entire sample is 400$\mu$m on a side, while the funnel barriers are each
27$\mu$m long, form a $60^{\circ}$ angle at their tips, and have a minimum
separation of 3.8$\mu$m between adjacent funnels.
If Brownian particles are placed at a uniform density in such a geometry, 
the density remains equal in both chambers of the sample.  In contrast,
when an equal number of swimming bacteria are introduced into each
chamber,
after roughly an hour there is a buildup of bacteria in the chamber that is on
the easy-flow side of the funnel barriers, as illustrated in Fig.~\ref{fig:2}(a,b).
In the figure, the easy-flow direction is to the right.
In Fig.~\ref{fig:2}(c), the
ratio $A$ of the bacterial concentration $\rho$ in
the right and left chambers, $A(t) = \rho_{R}(t)/\rho_{L}(t)$,
has an initial value of
$A=1.0$ at time $t=0$ and 
saturates to a value of $A=3.0$ at longer times, showing that a ratchet effect is occurring.
When the same experiment is performed with chambers separated by
symmetric holes instead of asymmetric funnels, 
then 
$A(t) = 1.0$ at all times, indicating that the asymmetry of the funnel is a crucial
ingredient for the occurrence of rectification.
When a mixture of swimming 
and non-swimming bacteria is
placed in the sample,
the swimming bacteria are rectified
while the non-swimming bacteria are not, showing that swimming is also important
in producing the ratchet effect.
Due to the complexity of bacterial systems, it was not 
obvious whether this initial observation of
an active matter ratchet effect could be understood as being produced solely by
the dynamics of the self-propulsion combined with
bacteria-wall interactions, or whether more complex bacterial behaviors play a role
in the ratcheting motion.

\begin{figure}[h]
\includegraphics[width=3in]{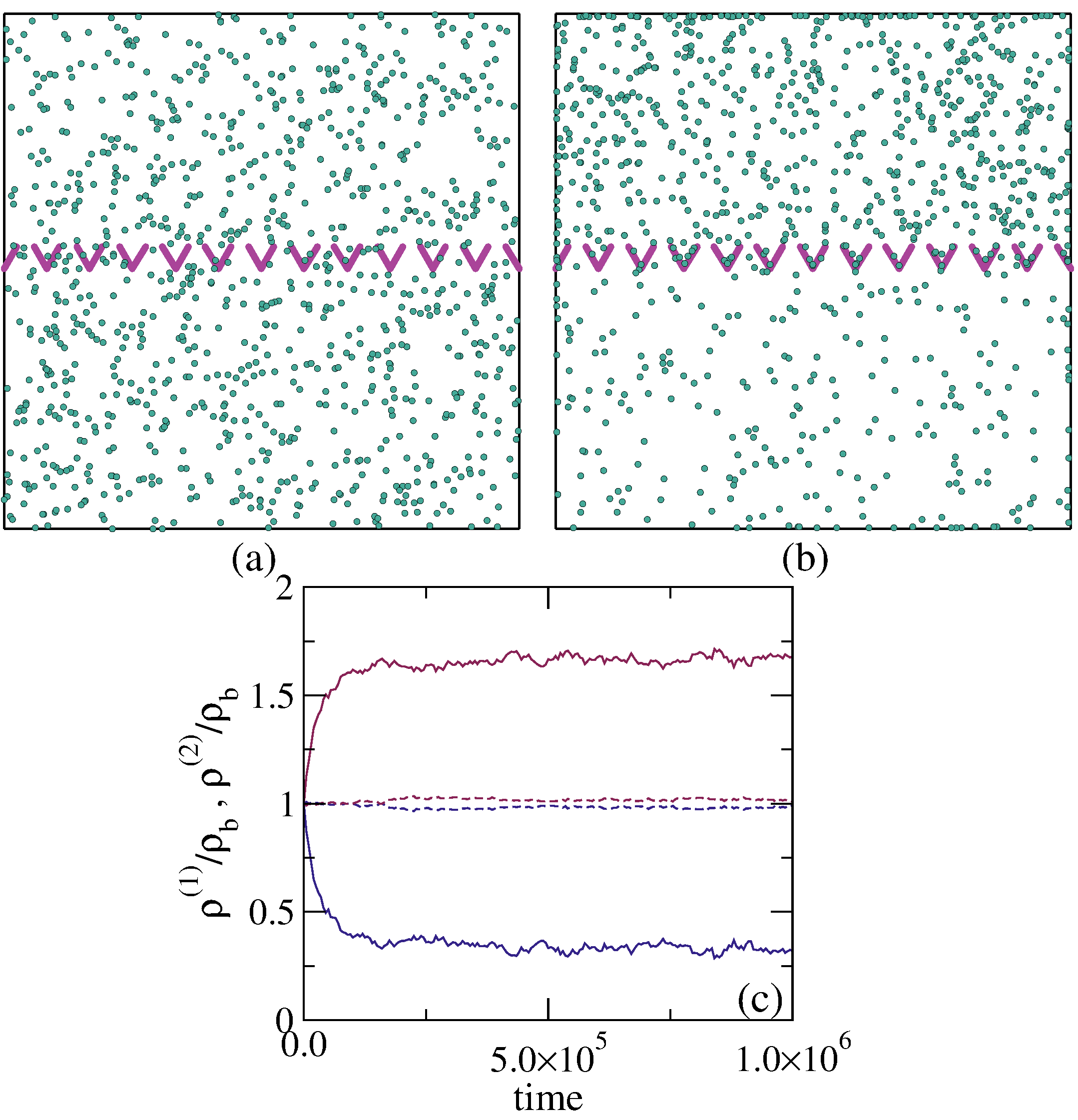}
\caption{ (a,b) Simulation images of noninteracting run-and-tumble
  particles moving between two chambers separated by a line of funnel-shaped barriers.
  The easy-flow direction of motion through the funnels is toward the top chamber.
  A particle striking a funnel or chamber wall runs along the wall until reaching the end
  of the wall or undergoing a tumbling event.
  (a) Initial uniformly dense condition
  for particles with a run length of $l_{b} = 40$ between tumbles.
  (b) After 100 tumbling events, the particle density in the upper chamber has increased.
  (c) Ratios $\rho^{(1)}/\rho_b$ (upper red line) and $\rho^{(2)}/\rho_b$
  (lower blue line) of the particle density
  $\rho^{(1)}$ in the upper chamber and $\rho^{(2)}$ in the lower chamber to the
  initial particle density $\rho_b$ as a function of time.
  Solid lines are for a sample with a run length of $l_b=180$ where there is a large
  rectification effect, and dashed lines are for a sample with a run length of
  $l_b=0.01$ where the rectification effect is negligible.
  Adapted from
M. B. Wan, C. J. Olson Reichhardt, Z. Nussinov, and C. Reichhardt,
Phys. Rev. Lett. {\bf 101}, 018102 (2008).
Copyright 2008 by the American Physical Society.
}
\label{fig:3}
\end{figure}

Wan {\it et al.} \cite{36} subsequently studied a simple model
of pointlike run-and-tumble particles moving
in a 2D container with the same funnel barrier geometry used in
the experiments of Ref.~\cite{7}, as illustrated in
Fig.~\ref{fig:3}.
In this model, particle $i$ obeys the following overdamped equation of motion:
\begin{equation}
\eta \frac{d {\bf R}_{i}}{dt} = {\bf F}^{m}_{i}(t) + {\bf F}^{T}_{i} + {\bf F}^{B}_{i} + {\bf F}^{S}_{i} 
\end{equation}
where $\eta$ is the damping coefficient and ${\bf R}_{i}$ is the location of
particle $i$.
Instead of the external driving force used in Eqn. 1, Eqn. 3 contains 
a motor force ${\bf F}^{m}_{i}(t)$ representing self-propulsion in a
randomly-chosen direction
that remains constant during
a running time $\tau$.
Every $\tau$ simulation time steps, 
particle $i$ undergoes an instantaneous tumbling event during which
the motor force is reoriented into a new random direction.  The particle
then runs in this new direction during the next time period $\tau$ before tumbling again.
The running speed $v$ is held constant.
For an isolated particle in the absence of any barriers, the motor force produces a motion
that is ballistic at short times and diffusive 
at very long times \cite{29}.
If the particle does not collide with any other objects,
then during a single running time it translates by one run length $l_b$, which is
a distance of $l_{b} = |{\bf F}^{m}_{i}|\tau$.
The stochastic thermal force ${\bf F}^{T}_{i}$ 
has the properties $\langle {\bf F}^{T}_{i}(t)\rangle = 0$ and
$\langle {\bf F}^{T}_{i}(t){\bf F}^{T}_j(t^{\prime})\rangle = 2\eta k_{B}T\delta_{ij}\delta(t - t^{\prime})$.
If ${\bf F}^{m}_{i}(t) = 0$ so that the particles experience only thermal fluctuations
but no motor force, then no rectification occurs.
${\bf F}^{S}_{i}$ represents particle-particle interactions, while
${\bf F}^{B}_{i}$ is the particle-barrier interaction term.
The barriers and confining walls are modeled as exerting 
a short-range steric repulsion on particle $i$, and a particle interacting with
a barrier or wall runs along the wall at a speed determined by the component of
${\bf F}^m_i$ that is parallel to the wall until the particle reaches the end of the wall
or undergoes a tumbling event that rotates its running direction away from the wall.

\begin{figure}[h]
\includegraphics[width=3in]{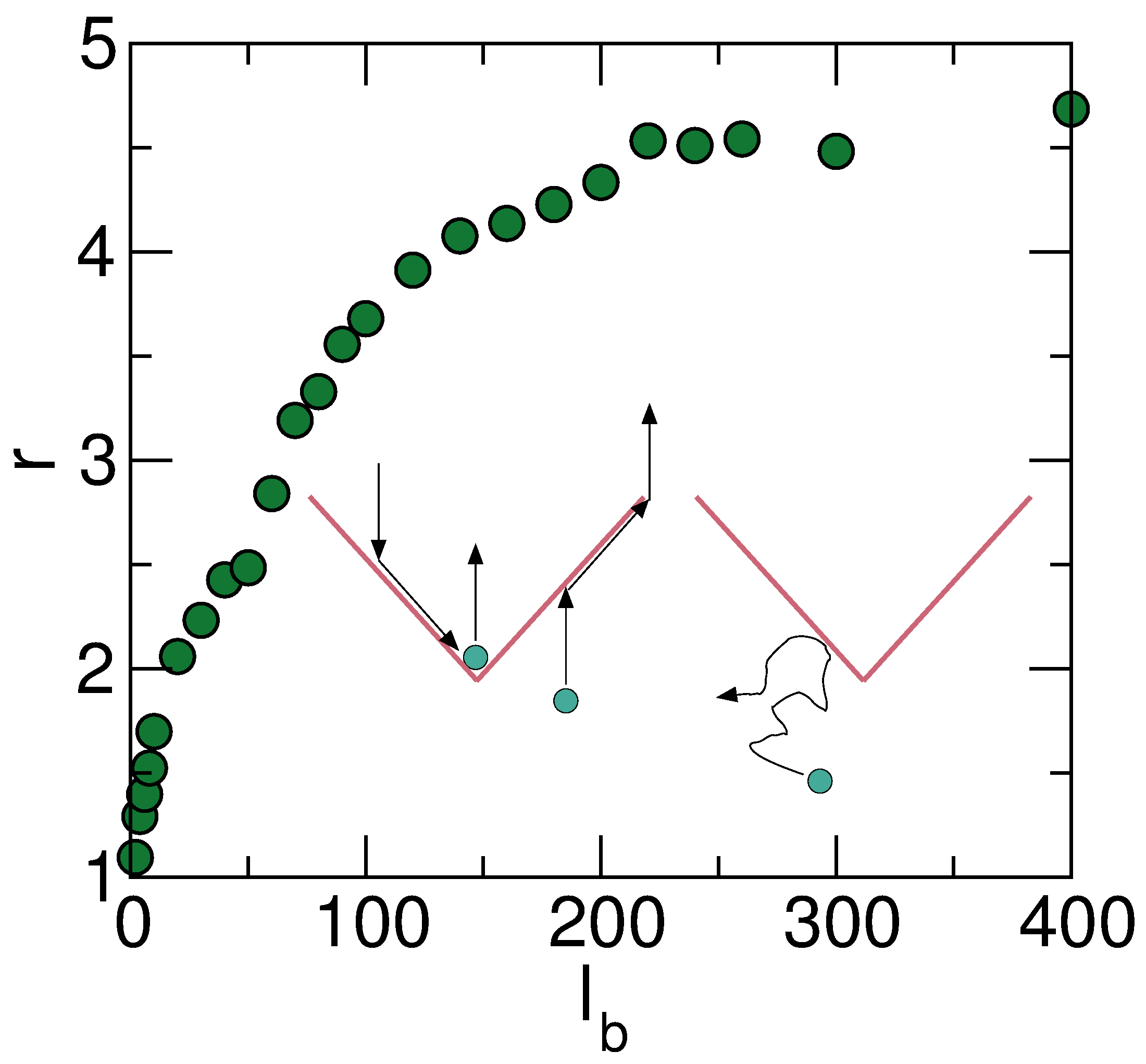}
\caption{The ratio $r=\rho^{(1)}/\rho^{(2)}$ of the
  particle density in the upper chamber to that in the lower chamber
  for the system in Fig.~\ref{fig:3} at steady state vs run length $l_b$, 
  showing that for small run lengths $r = 1.0$ indicating no rectification,
  while for large $l_{b}$ the ratio $r$ reaches a value of nearly $r=5.0$. 
  Inset:  Schematic illustration of the rectification mechanism.  On the left, an individual
  particle approaching the funnel barrier from above becomes trapped at the funnel
  tip, while a particle approaching from below is shunted by the funnel wall into the upper
  chamber.  On the right, a Brownian particle spends very little time in contact
  with the funnel walls.
  Adapted from
  M. B. Wan, C. J. Olson Reichhardt, Z. Nussinov, and C. Reichhardt,
Phys. Rev. Lett. {\bf 101}, 018102 (2008).  Copyright 2008 by the American Physical
Society.
}
\label{fig:4}
\end{figure}

A simulated system in which particle-particle interactions and thermal fluctuations are
both neglected, so that the particles experience only a motor force propulsion and
interactions with the barriers and walls, is illustrated
in Fig.~\ref{fig:3}(a,b).
The sample is of size $L=99a_0$ on a side, where $a_0$ is the unit of distance in the
simulation, and contains 12 funnel barriers that each have a tip angle of $60^{\circ}$ and an
arm length of $l_f=5a_0$.
The running length is
$l_{b} = 40a_0$, so a particle can move
much further than the length of a funnel arm during a single run time.   
The particle density is initially uniform, as shown in Fig.~\ref{fig:3}(a), but after
a time $100\tau$, Fig.~\ref{fig:3}(b) shows that the particle density has
increased in the upper chamber.
The ratio of the particle density $\rho^{(1)}$ in the upper chamber and
$\rho^{(2)}$ in the lower chamber to the initial particle density $\rho_b$ is plotted
as a function of simulation time steps in Fig.~\ref{fig:3}(c).
The solid lines show that in a system with $l_{b} = 180$, the density in the upper
chamber increases over time with a simultaneous decrease in the lower chamber density
until the sample reaches a steady state.
The dashed lines indicate that when the running length is very short, 
$l_{b} = 0.01$,
the density in both chambers remains nearly constant at a value equal to the
initial
particle density.
When $l_b$ is very small,
the particle motion is Brownian-like, so no rectification can occur.
The steady-state ratio $r = \rho^{(1)}/\rho^{(2)}$ is plotted as a function of
$l_{b}$ in Fig.~\ref{fig:4}.  For small $l_b$, $r=1$, while for large $l_b$, $r$ reaches
a value of nearly $r=5.0$.
The inset of Fig.~\ref{fig:4} illustrates the ratchet mechanism.
The left inset shows that
under the wall-following rule of motion, when $l_b$ is sufficiently large, a particle
approaching the funnel barrier from the top chamber becomes trapped at the funnel tip
until a tumbling event allows it to escape back into the top chamber, while a particle
approaching the funnel barrier from below is shunted along the barrier wall into the
top chamber.
In the right inset, a particle with a very short run time spends very little time interacting
with the wall, similar to a purely Brownian particle, and is unable to sample the
barrier asymmetry over a time scale long enough to bias its motion toward the upper
chamber.
The two ingredients that are required to produce a ratchet effect are the running of
particles along the walls and the breaking of detailed balance by the motor force.
In the same simulation study, the rectification effect is reduced when a finite
temperature is added since this increases the chance that a particle will prematurely
move away from a wall.
The ratchet effect is also reduced by the inclusion of steric particle-particle interactions,
since particle collisions effectively reduce the running length while a buildup of particles
in the funnel tip reduces the ability of the funnels to trap particles approaching from
the upper chamber.

\begin{figure}[h]
\includegraphics[width=3in]{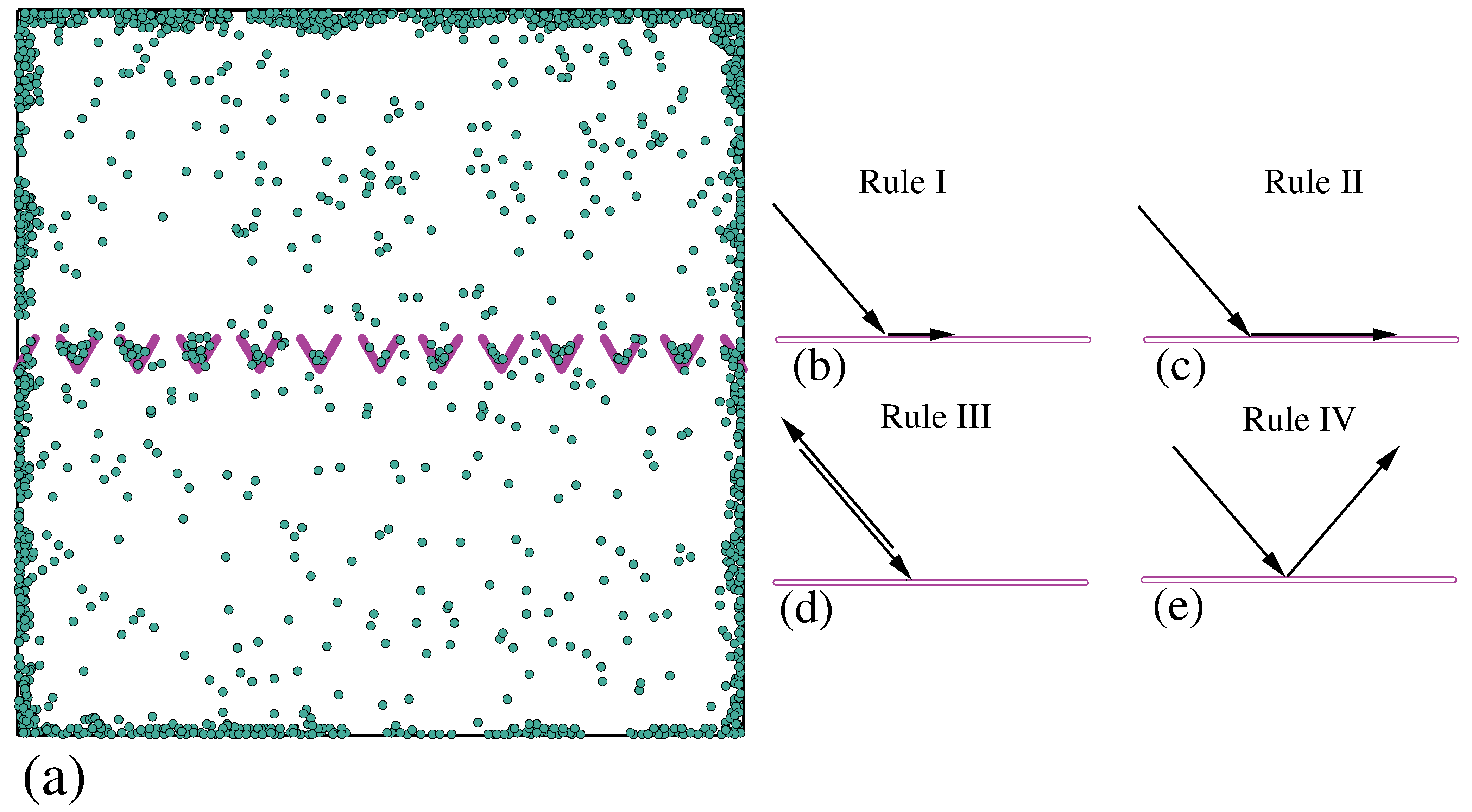}
\caption{(a) Simulation image of the buildup of sterically interacting run-and-tumble
  particles along the walls and in the funnel barrier tips for a system with a run length
  of $l_b=120$.
(b-e) Different possible particle-wall interaction rules.  (b)
Rule I: The particle maintains its
original running direction and moves along the wall with the component of its velocity
that is parallel to the wall.  (c) Rule II: The particle realigns its running direction to be parallel
with the wall. (d) Rule III: The particle reflects from the wall.  (e) Rule IV:
The particle undergoes an
elastic collision with the wall.  Ratchet effects occur for Rules I and II, with the strongest
ratcheting for Rule II, but Rules III and IV produce no ratcheting.
Adapted from
C.J. Olson Reichhardt, J. Drocco, T. Mai, M.B. Wan, and C. Reichhardt,
Proc. SPIE {\bf 8097}, 80970A (2011). 
}
\label{fig:5}  
\end{figure}

\begin{figure}[h]
\includegraphics[width=3in]{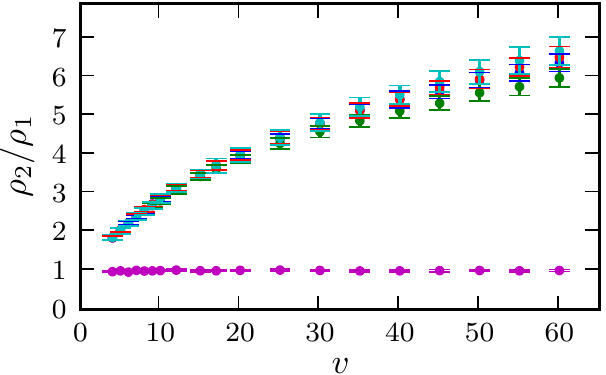}
\caption{
The simulated
rectification ratio $\rho_1/\rho_2$ for the particle density in chambers separated
by a line of funnel barriers for run-and-tumble
particles that align their running direction with any
wall they contact, showing increasing amounts of rectification for increasing
particle running speed $v$.
The purple curve with $\rho_{1}/\rho_{2} = 1.0$ is for the nonrectifying
case where the particles collide elastically with the walls.
Reprinted with permission from
J. Tailleur and M. E. Cates,
Europhys. Lett. {\bf 86}, 60002 (2009).
}
\label{fig:5b}  
\end{figure}

Figure~\ref{fig:5}(a) shows a snapshot from a simulation of run-and-tumble particles
with steric particle-particle interactions moving between two chambers separated by
funnel barriers.  When the run length is sufficiently large \cite{37}, 
the inclusion of steric interactions produces two additional features: a build up of particles
along the walls, and trapping of particles in the funnel tips.  Both of these effects were
observed in the experiments of Ref.~\cite{7}.
One question is whether the accumulation of particles along the walls in the experiments is
produced by hydrodynamic effects or whether it arises purely due to the self-propulsion of
the bacteria.
Using numerical and analytical methods,
Tailleur and Cates \cite{38} investigated run-and-tumble particles confined by a
spherical trap and found that the dynamics alone produce particle accumulation on
the walls.
Other studies of confined active matter where the motor force rotates diffusively
also show a build up of particle density along the walls,
and indicate that this
effect is enhanced near corners \cite{39,40}.
The tendency of active matter particles to accumulate
in corners is one of the
reasons the funnel barriers can produce
a ratchet effect, as the increased trapping occurs on only one side of the funnels
due to their curvature.
The work in Ref.~\cite{48} also shows how the
nature of the particle-wall interactions leads to rectification.
Rectification by the funnel geometry is robust against
changes in the running length distribution,  
as is observable by comparing the results
of Wan {\it et al.} \cite{36}, who used only a single running length in each simulation,
and of Tailleur and Cates \cite{38}, who considered a Poisson distribution of running
lengths.  
Changes in the nature of the particle-wall interactions, however, can destroy the
rectification.  Four possible interaction rules are shown schematically in
Fig.~\ref{fig:5}(b-e).  Under Rule I, the motor force alignment is unchanged by contact
with the wall and the particle moves according to the component of the motor force
vector that is parallel to the wall.  In Rule II, the motor force is realigned to be parallel
with the wall upon contact.  For Rule III, the particle is reflected by the wall, while in
Rule IV, the particle collides elastically with the wall.
In Ref.~\cite{38}, when the motor force realigning
Rule II is employed, efficient rectification of the particles
occurs; however, if the elastic collisions of Rule IV are substituted, the rectification is
lost, as shown in Fig.~\ref{fig:5b}.
An examination of all four wall rules in Ref.~\cite{37}
shows that the running of the particle along the wall, as in Rules I and II, is essential
for producing a ratchet effect.  When the particle does not remain in contact with
the wall, as in Rules III and IV, the rectification is lost.

\begin{figure}[h]
\includegraphics[width=2in]{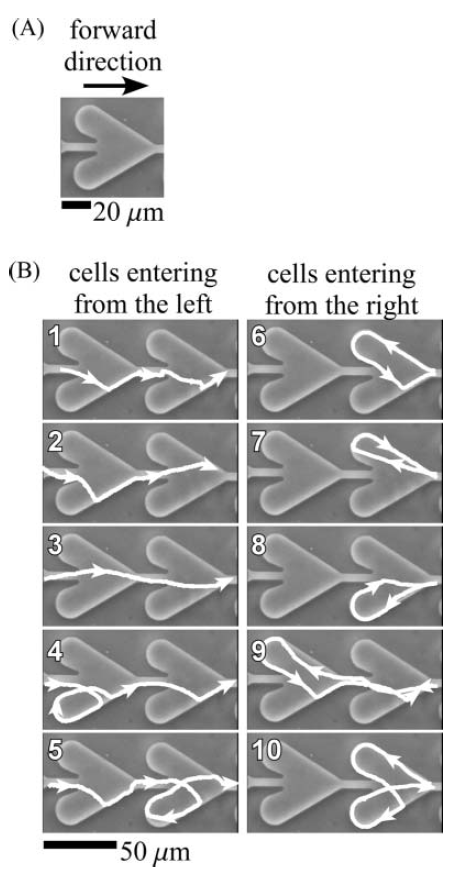}
\caption{The motion of bacteria through an array of asymmetric chambers.
  (A) The geometry of a single chamber, indicating the easy flow direction.
  (B) Experimentally obtained trajectories of bacteria.  Panels 1 through 5 show
  that bacteria can easily pass through the chambers for motion in the forward
  direction, while panels 6 through 10 show that motion in the reverse direction
  is blocked due to the chamber shape.
  Using this geometry 
  it is possible to fractionate swimming bacteria that have different sizes or
  mobilities.
  Reproduced from
  E. Hulme, W.R. DiLuzio, S.S. Shevkoplyas, L. Turner, M. Mayer, H.C. Berg, and G.M.
  Whitesides,
  Lab Chip. {\bf 8}, 1888 (2008), with permission of The Royal Society of Chemistry.
  http://dx.doi.org/10.1039/b809892a
}
\label{fig:6}
\end{figure}

Followup experiments by Galajda {\it et al.} \cite{41} using self-propelled 
large scale swimming bathtub toys produced a rectification effect
similar to that observed for the bacteria,
while experiments by Hulme {\it et al.} \cite{42} demonstrate
how the ratchet effect can be used to fractionate 
motile {\it E. coli} using arrays of asymmetric chambers
of the type illustrated in Fig.~\ref{fig:6}(a).
The arrows in panels 1 to 5 of Fig.~\ref{fig:6}(b)
show how bacteria moving in the positive $x$-direction are 
guided through the chamber array,
while in panels 6 to 10,
bacteria moving in the negative $x$ direction are blocked by the chamber geometry.
Variations on similar channel geometries have also been shown to produce guided motion
of bacteria \cite{43}.  
Kim {\it et al.} use similar microfabricated ratchet channels with asymmetry to 
concentrate {\it E. coli} at particular locations \cite{44}.  

\begin{figure}[h]
\includegraphics[width=3in]{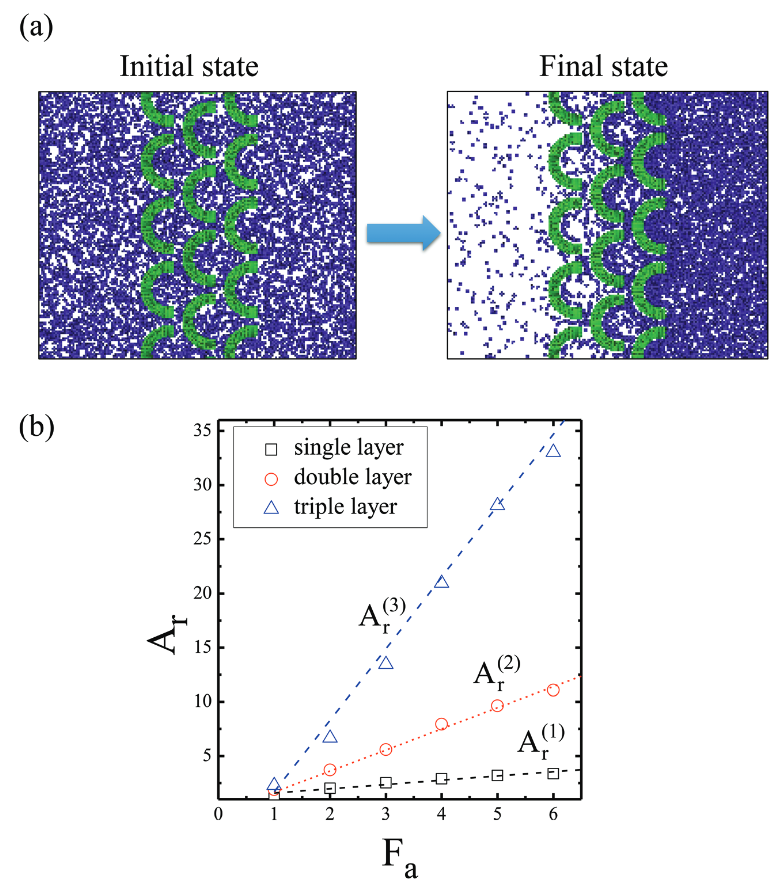}
\caption{ 
  (a)  Snapshots of the rectification effect in a simulation of particles with a motor force
  that rotates diffusively in
  two chambers separated by a triple layer of curved funnel barriers.  The initial state
  with a uniform particle density is shown on the left, while the final state with the
  particles rectified in the easy-flow direction into the right chamber is shown on the
  right.
  (b) The rectification ratio $A_r$, equal to the ratio of the density in the right chamber
  to the density in the left chamber,  vs motor force $F_{a}$ for samples containing
  one (squares), two (circles), and three (triangles) layers of funnel barriers.
  Reproduced from
  Y.-F. Chen, S. Xiao, H.-Y. Chen, Y.-J. Sheng, and H.-K. Tsao,
Nanoscale {\bf 7}, 16451 (2015), with permission of The Royal Society of Chemistry.
  http://dx.doi.org/10.1039/c5nr04124d
}
\label{fig:7}
\end{figure}

Numerous other shapes can produce rectification of bacteria, such as
the multiple rows of U-shaped funnel barriers illustrated in Fig.~\ref{fig:7}(a) that
produce rectification of particles with a motor force that rotates diffusively
\cite{45}.
In Fig.~\ref{fig:7}(b), a plot of the rectification ratio
$A_{r}$, given by the ratio of the density in the right chamber to that in the left chamber,
versus the motor force $F_{A}$ for one, two, and three rows of barriers shows that
$A_{r}$ increases linearly with $F_{A}$,
and that there is a strong enhancement of the ratchet effect
with the addition of more layers of funnels.
In the same study, an examination of barriers made from
open and closed triangles, U shapes, and half boxes indicates that
barriers with an open shape 
always produce a larger rectification than
barriers with closed shapes
due to the trapping of particles within the tips of the open barriers.

Theoretical and numerical studies show that
rectification effects are a general phenomenon that occurs
for self-driven particles in the presence of 
asymmetric piecewise periodic potentials \cite{46,New1}. 
In other numerical studies, the directed transport that occurs
for self-propelled particles interacting with
arrays of nonsymmetric convex obstacles without cavities
was shown to result from
the fact that the
particles tend to attach to solid surfaces \cite{47}. 

\subsection{Sorting in Funnel Arrays}
\begin{figure}[h]
  \begin{minipage}{3in}
   \hspace{1.0in}(a)\includegraphics[width=3in]{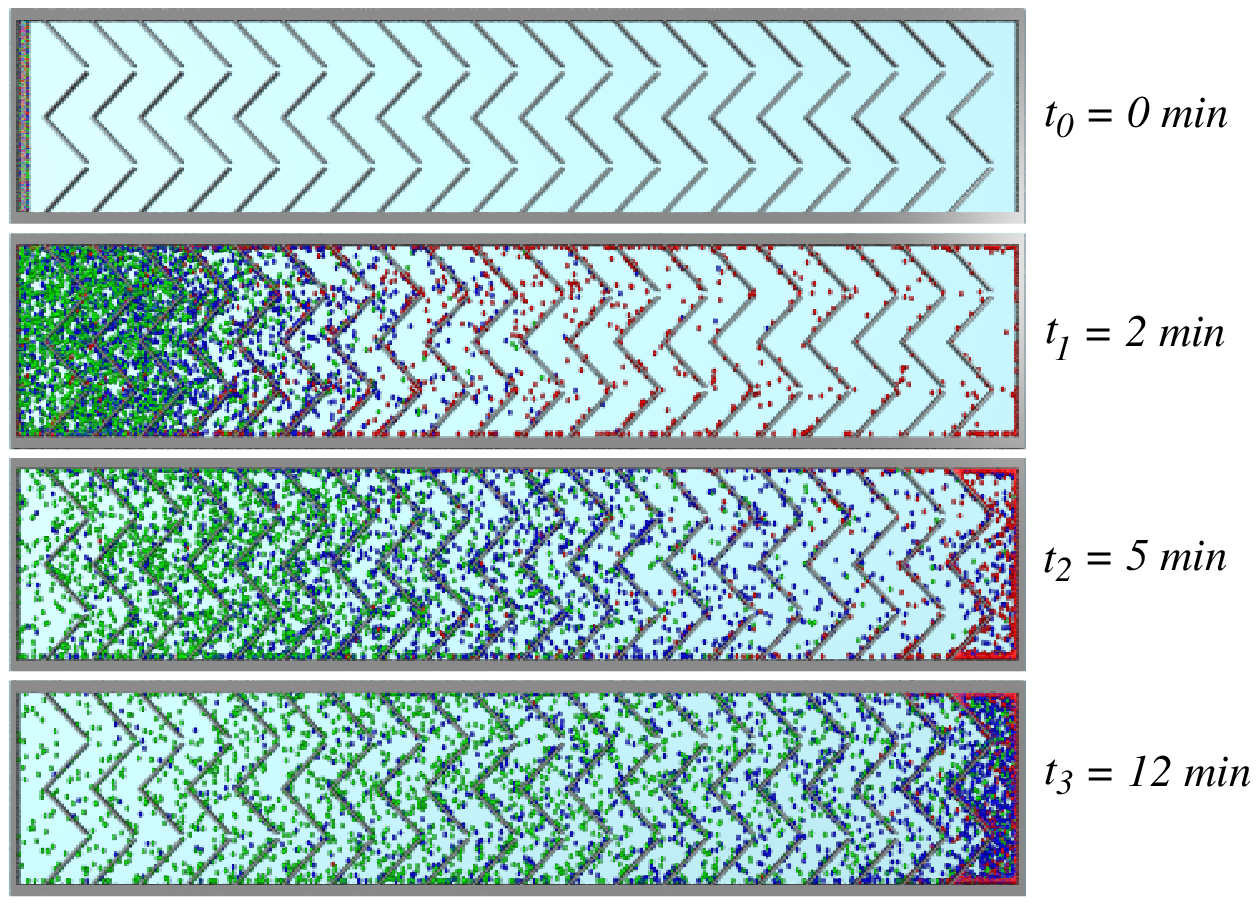}
    \end{minipage}\\
  \begin{minipage}{3in}
    \hspace{1.0in}(b)\includegraphics[width=3in]{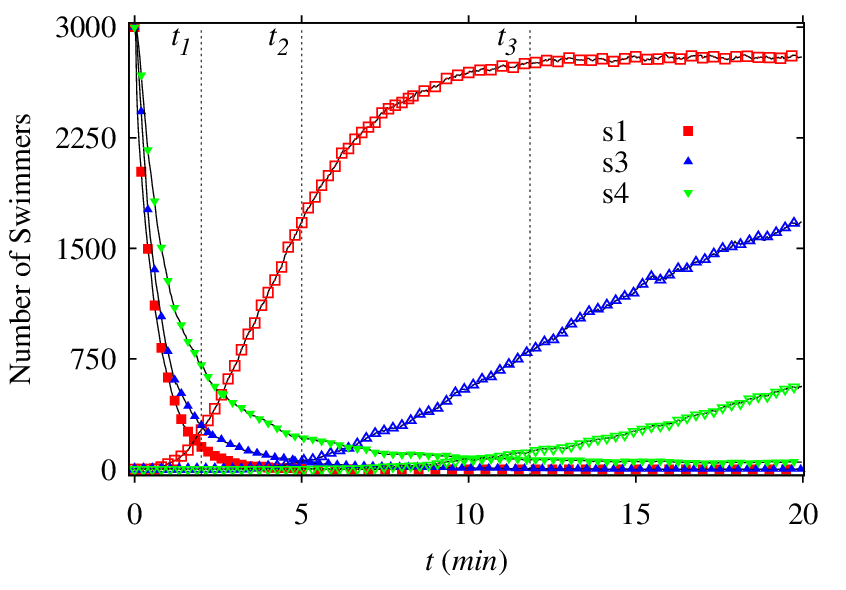}
    \end{minipage}
  \caption{(a) Snapshots of simulated
    bacteria positions at different times
    for a mixture of three {\it E. coli} mutants moving through a series of funnel barriers.
    Red: CheC497 or s1 (fastest), blue: AW405 or s3, green: Ucn602 or s4 (slowest).
    The bacteria begin along the leftmost wall at time $t=0$.
    (b) The number of bacteria vs time in the leftmost chamber (solid symbols) and the
    rightmost chamber (open symbols), showing that the different bacteria species can
    be fractionated using the funnel arrays.
    Reprinted with permission from
    I. Berdakin, Y. Jeyaram, V. V. Moshchalkov, L. Venken, S. Dierckx, S. J. Vanderleyden,
    A. V. Silhanek, C. A. Condat, and V. I. Marconi,
    Phys. Rev. E {\bf 87}, 052702 (2013).  Copyright 2013 by the American Physical
    Society.
}
\label{fig:8}
\end{figure}

In numerical studies of the motion of active particles with a diffusively rotating
motor force through funnel arrays,
I. Berdakin {\it et al.} examined the effect
of changing the swimming properties
by varying the distribution of run lengths, the magnitude of
the rotational diffusion, and the preservation of run orientation
memory after a tumbling event \cite{48}. They observe that there are optimal
swimming properties that maximize the rectification
as a function of the geometry of the funnels.
An illustration of the simulated funnel geometry appears in Fig.~\ref{fig:8}(a) along
with colored dots indicating the position of three different {\it E. coli} mutants modeled
on the basis of characterizations performed by Berg and Brown \cite{49}.
Red particles, or species s1, represent CHeC497 with a mean run duration of
$6.3$ seconds and a speed of 20 $\mu$m/s;
blue particles, or species s3, represent the wild type AW405
with a mean run duration of 0.86 seconds and a
speed of 14.2 $\mu$m/s;
and green particles, or species s4, represent Un602 which is the slowest with
a mean run duration of 0.42 seconds and a speed of 14.2 $\mu$m/s.
The particles are initially placed along the leftmost wall, 
and after 12 minutes the red particles, which have the longest 
run time and the highest speed,
have moved the furthest to the right,
followed by the blue particles, which move less far to the right, and
finally by the green particles, which have the least motion to the right. 
Figure~\ref{fig:8}(b) shows the number of bacteria as a function of time in the leftmost
chamber (decreasing curves) and rightmost chamber (increasing curves), indicating
that the different species become spatially separated over time.
In follow-up studies, the same group observed that the use of multiple arrays of funnels
makes it possible to sort species that have only a small variation in a single swimming
parameter \cite{50}.

\subsection{Eukaryotic Cells and Swimming Animals in Funnel Arrays}
The motion of biological cells through funnel geometries has been explored in several
experiments.
Kantsler {\it et al.} \cite{51}
considered swimming {\it Chlamydomonas} in both simulations and experiments
using four chambers separated by funnel barriers,
and measured the rectification ratio  $R = \langle N_{4}\rangle/\langle N_{1}\rangle$, where
$\langle N_{4}\rangle$ is the steady state number of cells in chamber $4$
and $\langle N_{1}\rangle$ is the steady state number of cells in chamber $1$.
For funnels with a tip angle of 35$^{\circ}$, $R=4.0$, indicating strong rectification.
In Ref.~\cite{52}, experiments on human sperm cells
in quasi-2D microchambers containing funnel barriers show that rectification occurs
when the cells become trapped in the funnel tips.  The same study showed that since
the sperm are attracted to the boundaries, the ratchet effect can be optimized by
using U-shaped barriers.
Experiments with {\it C. elegans} moving through funnel barriers showed
that the wild type, which moves along walls that it contacts, can be rectified by
the barriers \cite{53}.  In contrast, mutant strains that lack touch sensory neurons in their
body reverse motion upon contacting a wall instead of following the wall, and these
mutants are not rectified by the barriers.

\subsection{Other Active Ratchet Geometries}
A variety of other ratchet geometries for active particles have been studied, including
2D corrugated asymmetric substrates \cite{N6} and
2D geometries where entropic effects are important \cite{N7}.  Other works
focus on
the effect of changing the active particle shape to
swimming ellipses \cite{N8}, or
on including an additional external drift force to the active particles to produce
an active drift ratchet \cite{N9}.
Kulic {\it et al.} proposed that ratchet effects can  
occur in a variety of plants where the asymmetry of the roots, seeds or
leaves is important for transport 
in the natural world where fluctuations may be present \cite{N10}.
Finally, in the case of symmetric substrates, it has been proposed that by periodically
modulating the velocity of the active particles and then phase-shifting this velocity
with respect to the substrate periodicity, enough symmetries are broken to permit
a ratchet effect to occur \cite{N11}.

\subsection{Future Directions}
In ratchets produced using funnel barriers, it would be interesting to
examine additional modes of motion of the active particles 
such as particles that avoid walls, 
particles that reverse their motion if the density is too high,
or particles that change swimming strategies as a function of local density or of time.
Many of the active matter ratchet studies
performed so far employ infinitely repulsive barriers or walls;
however, many previous studies of nonactive matter ratchet effects produced by
external driving were performed for substrates containing pinning sites with a finite
strength, so that a particle experiencing a sufficiently large driving force can pass
through the pinning potential or barrier.
It would be interesting to examine active matter particles moving through
asymmetric finite-strength pinning arrays rather than obstacle arrays.
Pinning arrays for colloidal particles can be readily created experimentally using
optical means, so it should be possible to create optically generated pinning arrays
for active colloidal particles or swimming bacteria.
Additionally, optical traps introduce the possibility of creating
flashing substrates, which could significantly enhance active matter ratchet effects.    
It was recently proposed that by using a light intensity pattern with a funnel or
chevron shape to manipulate a dense assembly of sterically interacting
active particles that are attracted by light, the patterned regions would become filled
with trapped active particles that would then serve as an asymmetric substrate
that can produce ratcheting motion of the remaining untrapped active particles \cite{54}.
There has also been work 
examining how asymmetry in the curvature of
the walls can produce different pressures on each side of the wall,
producing a ratchet effect \cite{55}.
This suggests that curvature is another route
to explore in producing different types of active ratchet devices.
Other possible systems to explore include crawling insects or animals. 
Studies that reveal boundary following behaviors in animals \cite{56}
suggest that asymmetry in the boundaries
could be used to produce ratchet effects.
Another avenue of study is the introduction of feedback effects to a ratchet system,
which could be achieved through optical methods.
Recent studies 
of active matter using feedback controls
have produced a variety of behavior for very simple feedback rules \cite{New4,New5}.  

\section{COLLECTIVE ACTIVE MATTER EFFECTS}

\subsection{Active Ratchet Reversals}
\begin{figure}[h]
\includegraphics[width=3in]{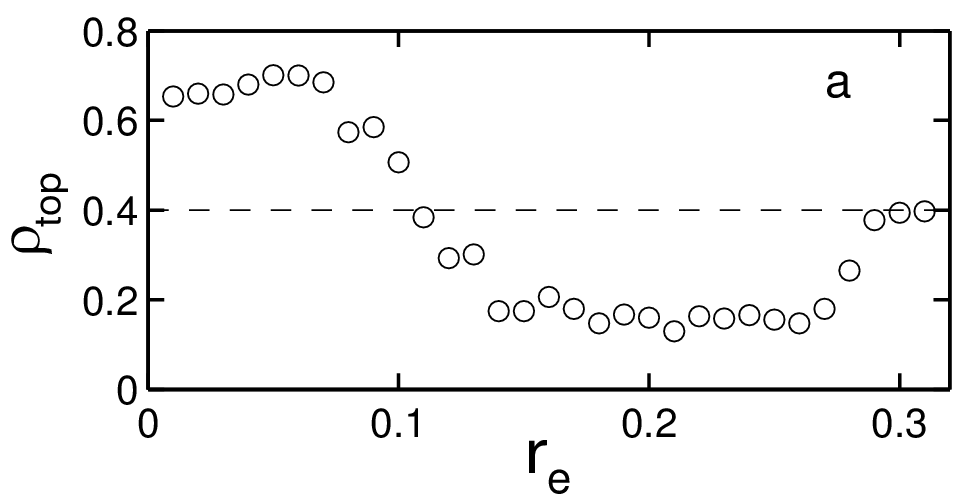}
\includegraphics[width=3in]{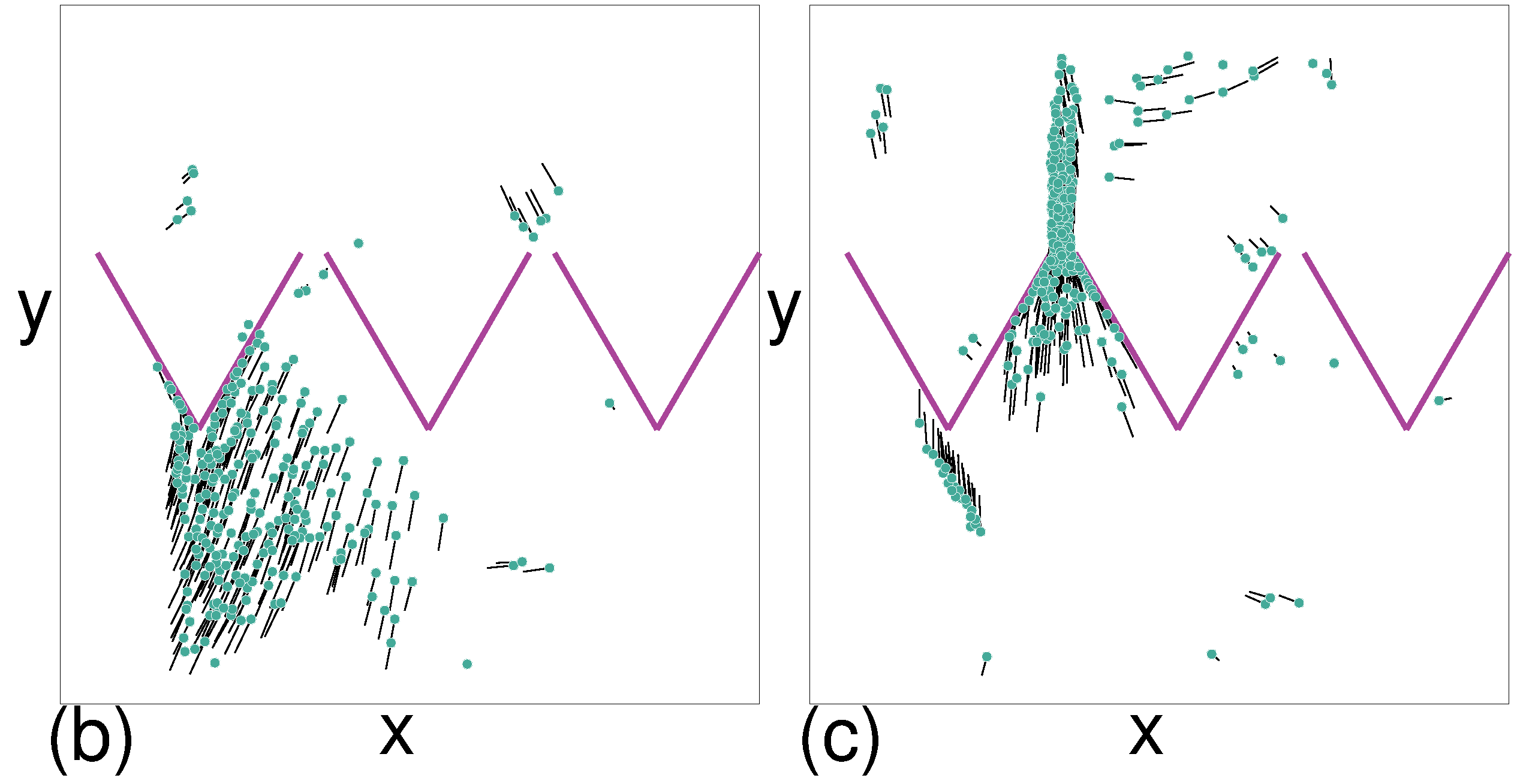}
\caption{Simulations of sterically interacting flocking particles in a funnel barrier array.
  (a) Steady state density $\rho_{\rm top}$ of particles in the upper chamber
  vs the particle exclusion radius $r_e$.  The dashed line indicates the uniform starting
  density with $\rho_{\rm top}=\rho_{\rm bottom}=0.4$.
  The ratchet effect in the easy flow direction is maximized for small $r_e$.  A reversal
  in the ratchet direction occurs
for $0.1 < r_{e} < 0.3$, and the ratchet effect is lost
for $r_{e} > 0.3$ when the particles become too large to fit between adjacent funnels.
(b,c) Images of a small portion of the sample illustrating the
forward rectification mechanism.
(b) A flock approaches the funnel barriers from below.
(c) The flock elongates and passes through the barrier array as a unit.
Adapted from J.A. Drocco, C.J. Olson Reichhardt, and C. Reichhardt,
Phys. Rev. E {\bf 85}, 056102 (2012). Copyright 2012 by the American Physical
Society.
}
\label{fig:9}
\end{figure}

\begin{figure}[h]
  \begin{minipage}{2in}
    \includegraphics[width=2in]{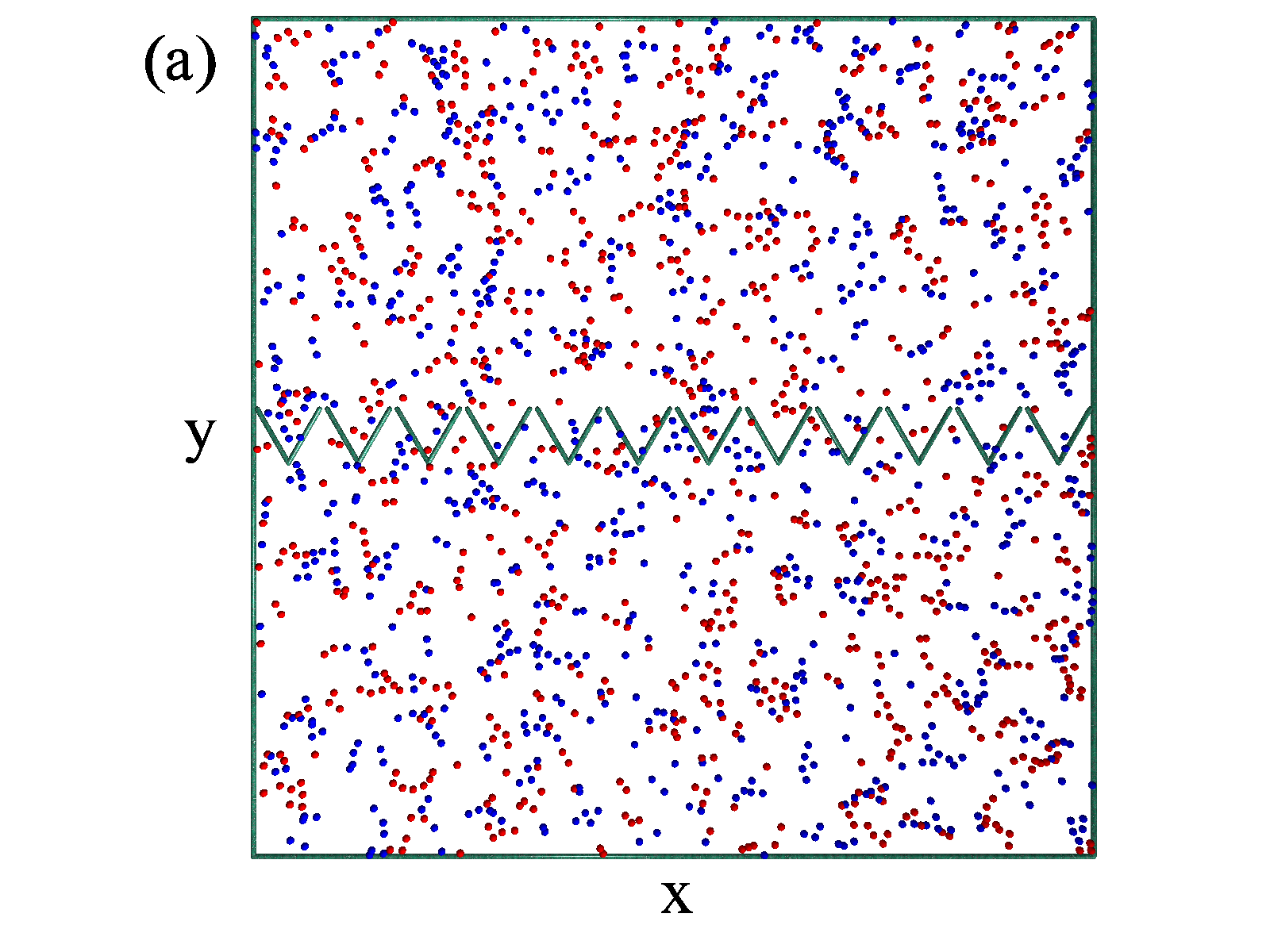}
  \end{minipage}%
  \begin{minipage}{2in}
    \includegraphics[width=2in]{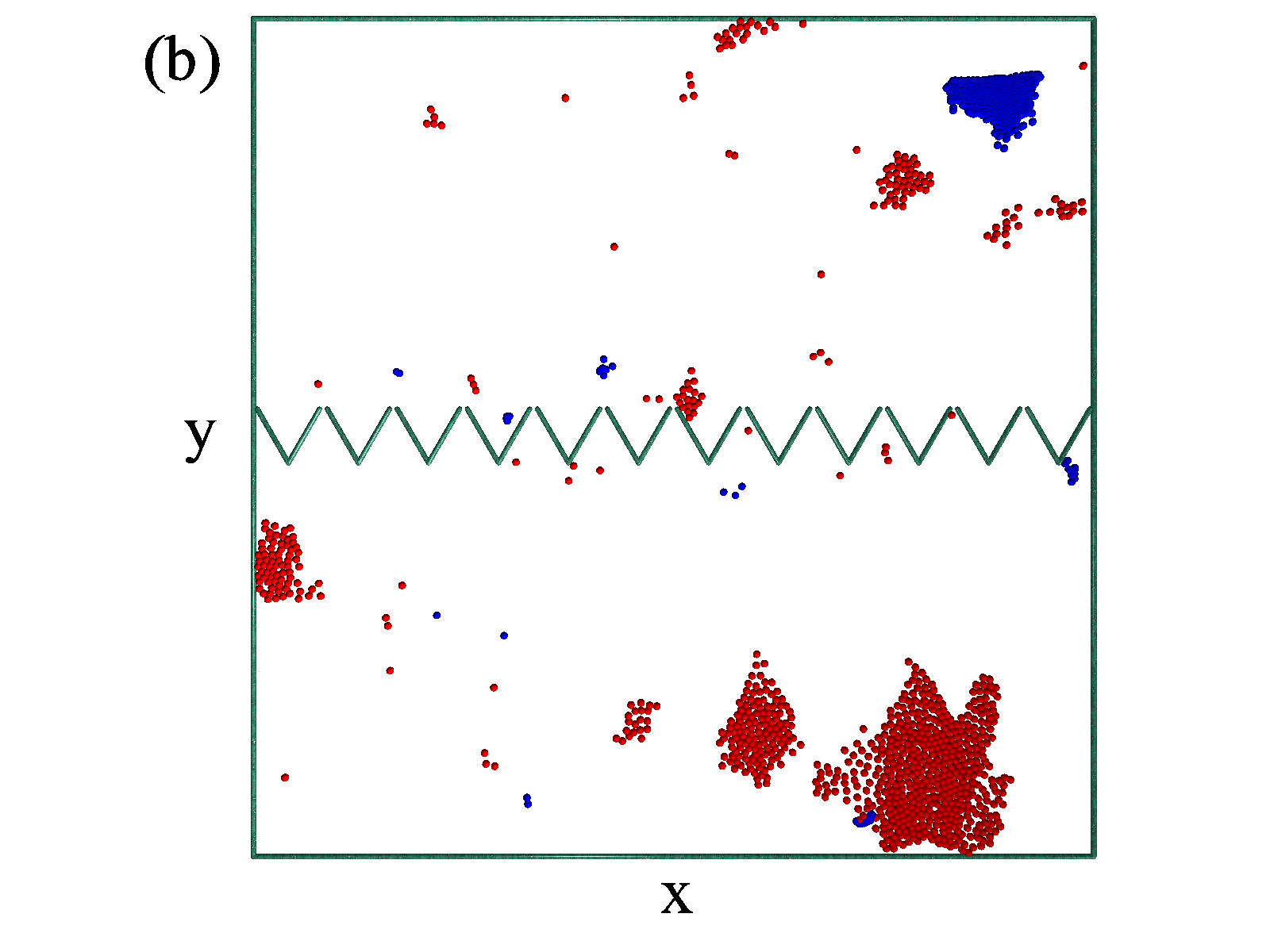}
    \end{minipage}
  \caption{Images from simulations of sterically interacting
    flocking particles with two different
    exclusion radii, $r_e=0.055$ (small, blue) and $r_e=0.22$ (large, red).
    (a) The initial configuration with uniform density in both top and bottom chambers,
    which are separated by a line of funnel barriers.
    (b) A representative steady state configuration at later time.
    The larger red particles undergo reversed rectification and collect in the bottom
    chamber, while the smaller blue particles undergo forward rectification and collect in
    the top chamber.
}
\label{fig:10}
\end{figure}

In rocking and flashing ratchet systems, a rich variety of collective effects
including ratchet reversals 
arise when particle-particle interactions are included \cite{2,3,4,5}. 
For run-and-tumble active particles moving through funnel barriers,
Wan {\it et al.} observed that the addition of steric
particle-particle interactions produced only a monotonic reduction in the
rectification effect \cite{36}.
Drocco {\it et al.} \cite{57} placed particles obeying
a Vicsek model, modified to include particle-barrier interactions and
steric particle-particle interactions with an exclusion radius $r_e$,
in a funnel barrier geometry.
For  small $r_{e}$, the particles ratchet in the easy flow direction,
and the magnitude of the ratchet effect depends
on the noise term $\eta$ in the Vicsek model.  For small $\eta$ the system forms
flocks, while for large $\eta$ the system is in a liquidlike state.
In the funnel barrier system, 
the ratchet effect is strongly reduced or completely lost for large $\eta$, while
for small $\eta$ where flocking occurs, a pronounced ratchet 
effect appears with density ratios in the two chambers approaching
$r=5.0$.
Figure~\ref{fig:9}(b,c) highlights the enhancement of the ratchet effect due to the
collective rectification of groups of active particles.
In Fig.~\ref{fig:9}(b), a flock approaches the funnel barriers from below.  The flock
becomes compressed and moves through the barrier array as a unit, as illustrated
in Fig.~\ref{fig:9}(c).
A reversal of the ratchet effect, in which the particles collect in the bottom chamber
in the hard flow direction of the funnel barriers, can occur when flocks are
present.
Figure~\ref{fig:9}(a) shows a plot of the particle density in the top chamber
$\rho_{\rm top}$ versus
the particle exclusion radius $r_e$ for a system in which
$\eta=1.5$, the flocking radius is 1.0, the starting uniform density
is $\rho=0.4$, and the closest spacing between adjacent funnel barriers
is $l_0=0.6$.
For $0.0 < r_{e} < 0.1$, rectification occurs in the easy flow direction of the funnel
barriers as illustrated in Fig.~\ref{fig:9}(b,c), but there is a reversal of the rectification
into the hard flow direction for
$ 0.1 \leq r_{e} < 0.3$.
For $r_{e} \geq 0.3$, the diameter of the particles is  larger than the
closest spacing between funnels, $r_e\geq l_0$, so it is no longer possible for particles
to pass between the upper and lower chambers, and the rectification effect
is lost.
The ratchet reversal is produced by a jamming effect that occurs during the compression
of a flock of particles as it approaches the opening between adjacent funnel barriers
from below.
The compression causes the particles at the leading edge of the flock to form a rigid
close-packed structure that is wider than the interfunnel spacing $l_0$, and makes
it impossible for the flock to cross the line of funnel barriers.
In contrast, when a flock approaches the funnel barriers from above, the barriers cleave
the flock into smaller clusters, and during this process one or two particles can escape
from the flock and pass between the funnel barriers to enter the lower chamber.  This
reversed motion process always occurs but cannot compete with the forward motion
of flocks of particles moving into the top chamber for small $r_e$.  When the forward
motion is inhibited by jamming at larger $r_e$, the reversed motion dominates and
a reversed ratchet effect appears.
The reverse ratchet effect operates much more slowly
than the forward ratchet, so a much longer time
is required for the system to reach a steady state in the reverse ratchet regime than in
the forward ratchet regime.
If two species of particles with different exclusion radii of
$r_e  = 0.055$ (small) and $r_{e} = 0.22$ (large)
are placed in the container, flocks of the different species ratchet in opposite directions
and can be spatially separated, as shown in Fig.~\ref{fig:10}.

Ratchet reversals have also been observed for eukaryotic cells moving on
asymmetric substrates.
Mahmud {\it et al.} \cite{58} examined cell motility on micropatterned
ratchet surfaces.
They considered cancerous B16F1 cells, which are a mouse epithelial-like melanoma cell,
non-cancerous Rat2 fibroblast cells,
and cancerous MDA-MB-231 human mammary gland cells.
For the B16F1 and Rat2 cells, directed motion occurs along
the easy flow direction of the funnel array.  The addition of a protruding spike shape
to the funnel shaped chambers does not affect the Rat2 cells, which still
ratchet in the easy direction, but causes the B16F1 and MDA-MB-231 cells to perform
a reversed ratchet motion in the hard flow direction.
The differences in the direction of the ratchet motion for the different cells were attributed
to the different crawling mechanisms of the cells.
Rat2 cells use long protrusions that can grab the funnel edges or the ends of the spike
shapes, while the other cells prefer to spread out in the open spaces as they move.
Mahmud {\it et al.}
also found that it is possible to use arrays of the spike shaped funnels to separate a mixture
of different cells over time,
and they propose the creation of a bidirectional ratchet, opening the possibility of
creating cell traps and ratchets to achieve various types of medical treatments.
Other studies have shown that
unidirectional cell motion can be achieved using asymmetric nanotopography \cite{59}.

\subsection{Active and Passive Ratchet Mixtures}

Self-propelled Janus colloids, whose direction of motion slowly diffuses
rotationally, represent another type of experimental artificial active matter system.
Gosh {\it et al.} \cite{60} performed simulations 
of active Janus particles in an asymmetric channel,
and observed that rectification occurs in the easy flow direction of the channel and that
the magnitude of the ratchet effect increases with increasing correlation time of
the particle running direction, eventually saturating for large correlation times.
They also  considered a mixture of $N_{m}$ active particles
and $N_{p}$ non-active particles in the same geometry,
with the particle-particle interactions modeled
as a repulsive harmonic force, and found that the active 
particles can induce 
a net rectification of the non-active particles.
A particularly interesting effect is that the rectification efficiency for the non-active
particles is non-monotonic as a function of particle density.  For one choice of parameters,
the maximum non-active particle ratchet efficiency occurs for a ratio of
$N_p/N_m=12$, indicating that even a very small fraction of active particles can
efficiently rectify the passive particles.
In the dense limit of $\phi = 1.0$ close to the jamming regime, Ghosh {\it et al.} find
a small reversal of the ratchet effect.    
 
\begin{figure}[h]
\includegraphics[width=3in]{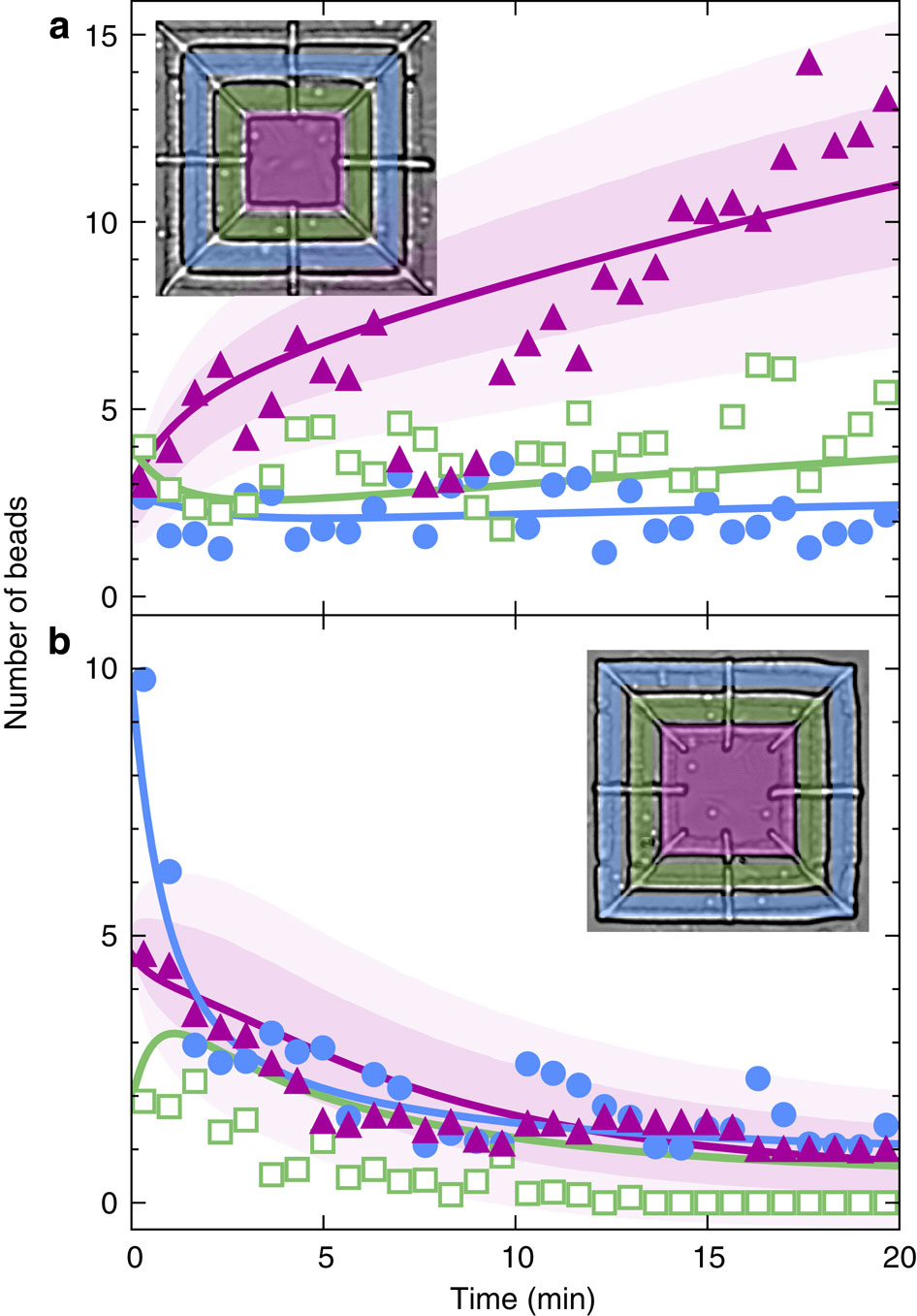}
\caption{ Experimental demonstration of directed motion of
  passive colloids immersed in a bath of bacteria which interact with a
  ratchet geometry.
  The plots indicate the number of passive particles in each chamber or spatial region, color
  coded according to the inset.
  (a) A geometry that concentrates the passive particles at the center of the device.
  (b) A geometry that ejects the passive particles.
Reprinted by permission from MacMillan Publishers Ltd: Nature Communications,
N. Koumakis, A. Lepore, C. Maggi, and R. Di Leonardo,
Nature Commun. {\bf 4}, 2588 (2013), copyright 2013.
}
\label{fig:11}
\end{figure}

In experiments with passive colloids immersed in a bath of swimming bacteria, 
Koumakis {\it et al.} showed that introducing an asymmetric substrate that causes
the bacteria to undergo rectification also produces transport of the passive colloids
into or out of enclosed regions
\cite{61}. 
Figure~\ref{fig:11}(a) shows the time evolution of the number of passive colloids in
each chamber of a sample containing three layers of ratchets for a geometry that
concentrates the colloids in the central chamber, while Fig.~\ref{fig:11}(b) shows
the same quantity for a geometry that ejects the colloids from the box.
Koumakis {\it et al.} also found that 
curved ratchet geometries were not as efficient 
as flat boundaries in their device since the colloids would follow curved trajectories
over the curved surfaces.
These results show how collective effects in active ratchets could
be used to achieve a variety of active pumping techniques.
It may also be possible that biological systems such as cells already
take advantage of such collective ratchet effects.

\subsection{Future Directions}

The field of collective effects in active matter systems is wide open.
Since a wide variety of flocking models exist,
it would be interesting to determine what
types of ratchet effects occur for 
different flocking rules.
This would also be an interesting question to examine in experiments on schools of fish
or swarming insects which could be placed in funnel like channels, with a comparison
between ratchet behaviors for schooling and non-schooling systems.
Similar experiments could be performed with herding animals.
It would also be interesting to explore systems with competing effects, such as
a ratchet geometry that biases the motion of the system in one direction
combined with some other
competing force that biases the motion of the system in the opposite direction.
In such a case, flocking or collective effects could
be turned on or off to take advantage of or nullify the ratchet effect. There 
has already been some evidence for competing interactions that can overwhelm
the ratchet effect, such as an experiment on bacteria in a funnel barrier ratchet
geometry,
where the bacteria form Fisher wave
swarms
that move against the ratchet direction \cite{62}. 

Self-driven colloids have been shown to exhibit a phase separation transition into gaslike
and clustered solidlike phases for high enough 
density or activity \cite{31,32,33,34},
so it would be interesting to examine how the ratchet efficiency is affected by
the onset of self-clustering.
Introduction of a substrate can modify the density or activity at which the self-clustering
begins to occur, 
as well as the mean distance that individual particles can move \cite{63}. 
Collective ratchet effects should also depend strongly on the type of interactions
between the particles.
In the systems studied so far, short-range steric repulsion has been used for the
particle-particle interactions; however, it is also possible to consider active
particles with
Lennard-Jones interactions, long-range repulsive interactions,
competing long-range repulsion and short-range attraction, or
many body interactions.

Another class of system is active membranes or sheets
that could mimic cells or collections of tightly bound
organisms.
Swimming deformable droplets are one example of
an extended active object of this type \cite{New3}, and if such droplets were placed
on an asymmetric surface, ratchet effects could occur.
In this case, varying the substrate periodicity to be larger or
smaller than the size scale of the active object could cause changes in the
ratchet effect.  There has already been a study demonstrating ratchet effects for
active polymers in funnel barrier arrays \cite{64}.
Other motion rules could also be introduced
to model objects that have mechanics similar to cells, such
as protrusions that push or pull to mimic the experiments in Ref.~\cite{58}.
There could also be ratchet effects for extended active objects that can expand
or contract, such as a squirmer model or a system with inertial dynamics.

Collective ratchet effects can 
also be examined for groups of interacting rods coupled by molecular motors
when an additional asymmetry is introduced, or active gel systems such
as active nematics \cite{65,66} could be studied in the presence of an asymmetric
substrate.
Active nematics can also exhibit 
mobile defects \cite{67,68}, so it may be possible
to produce ratcheting motion of the defects.

\section{RATCHETS IN CIRCLE SWIMMERS AND CHIRAL SYSTEMS} 

There are many examples of active biological and artificial systems
that exhibit a chirality in their motion,
such as biological circle swimmers \cite{69,70,71} and artificial
circle swimmers \cite{72,73,74}.
In this case, ratchet effects can occur even in the absence of
a substrate asymmetry since the chirality of the swimming introduces an
asymmetry to the system.
One of the first examples of a ratchet effect produced by chirality was observed for
circularly moving particles on a  
2D periodic substrate \cite{75}.  As the substrate-free particle orbits are varied
from circular to
more complex Lissajous patterns, a series of transitions occurs between
localized closed trajectories with no net translation to open trajectories that permit the
particles to translate in a particular direction, producing a ratchet effect. 
Additional simulations have shown
that directed motion can be achieved for circularly moving particles on periodic
substrates, where various types of translating trajectories appear \cite{76},
while in experiments with circularly moving colloids on 2D substrates, directed
or ratchet motion can be produced \cite{77}.
Nourhani {\it et al.} \cite{78} considered a model of
counterclockwise circle swimmers on a 2D periodic substrate for both
deterministic
and stochastic swimmers.
In the deterministic regime they found that the particles can either 
form closed orbits or translate at fixed angles with respect to the substrate periodicity. 
In the stochastic regime, the ratchet effect extends
over a wider range of parameters since the particles can hop between
different rectifying orbits even if they become temporarily trapped in a
closed or nonrectifying orbit.

\begin{figure}[h]
\includegraphics[width=3in]{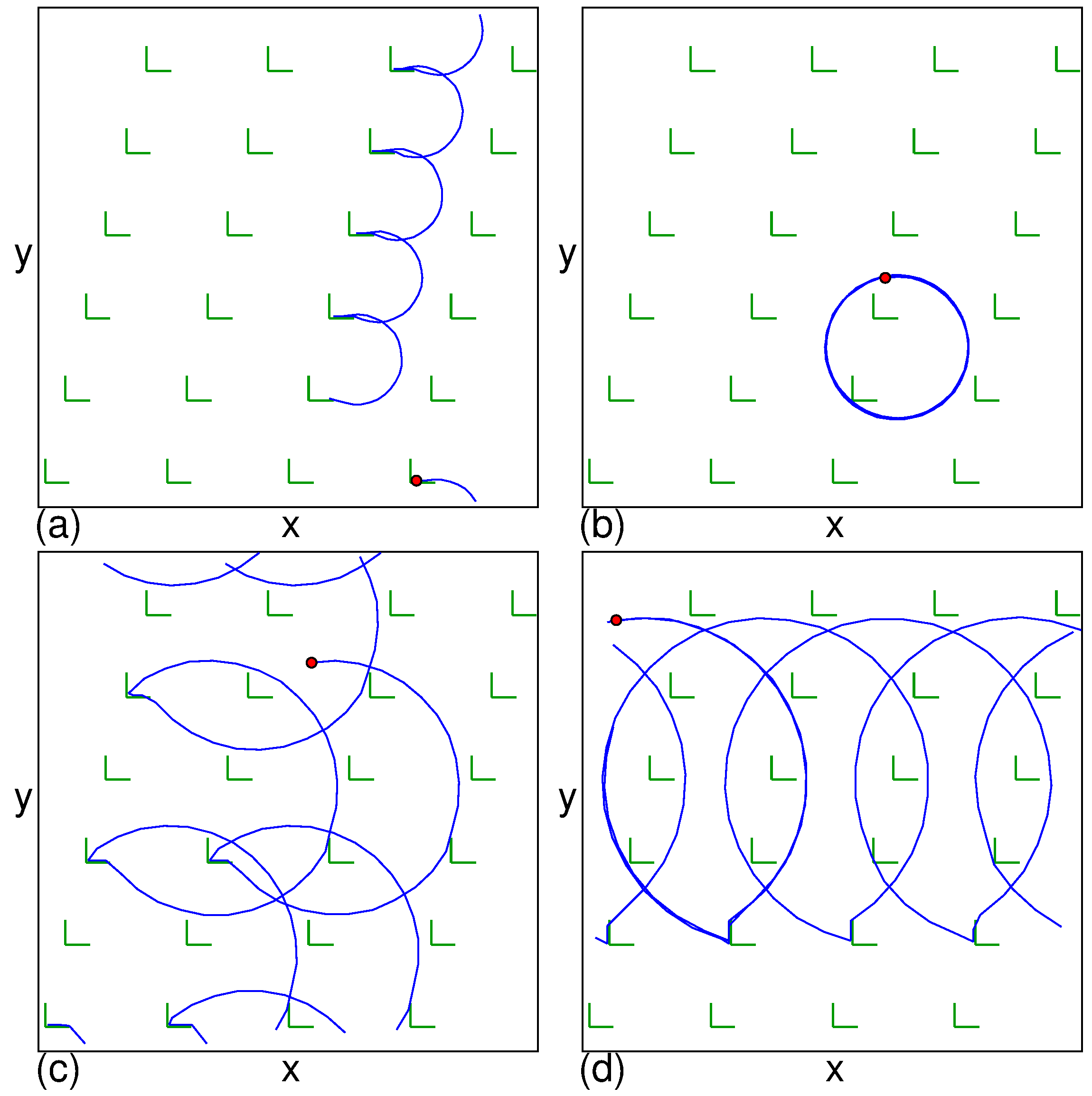}
\caption{ The particle orbits from a simulation of circle swimmers
  in an array of L-shaped barriers.
(a) At a motor force of $A = 0.57$, translation occurs in the positive $y$ direction. 
  (b) For $A = 0.9$ a non-ratcheting orbit appears.
  (c) At $A = 1.6$ the particle translates in the positive $x$ and $y$-directions. 
(d) At $A = 2.05$ the particle translates in the negative $x$-direction. 
  Adapted from
C. Reichhardt and C. J. Olson Reichhardt,
Phys. Rev. E {\bf 88}, 042306 (2013). Copyright 2013 by the American Physical
Society.
}
\label{fig:12}
\end{figure}

\begin{figure}[h]
\includegraphics[width=3in]{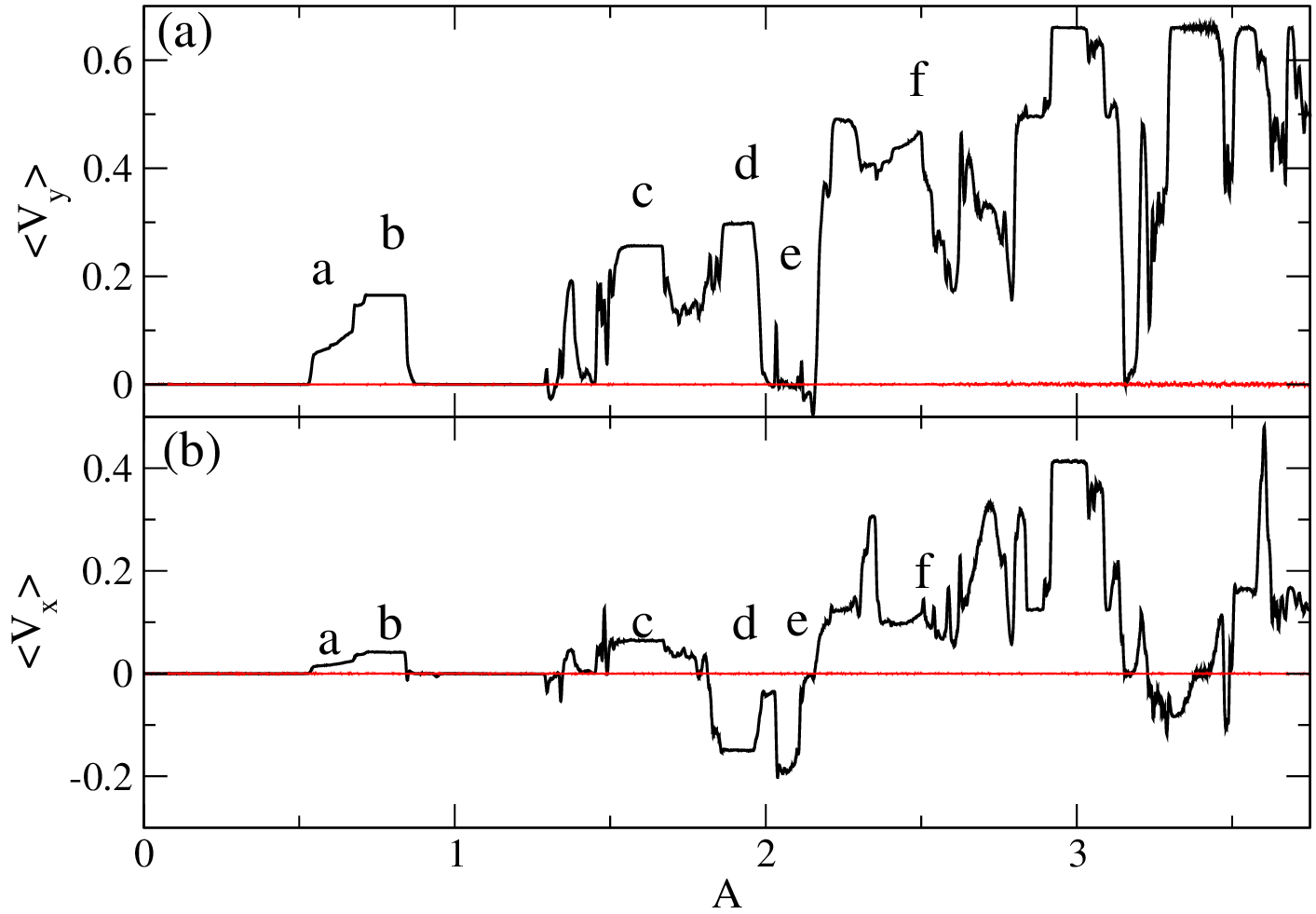}
\caption{ The net velocity of the circularly swimming particle from the system
  in Fig.~\ref{fig:12} showing a series of ratchet orbits as a function of the motor
  force amplitude $A$.
(a) $\langle V_{y}\rangle$ vs $A$. (b) $\langle V_{x}\rangle$ vs $A$. 
  If the particle swimming direction is reversed from counterclockwise to clockwise,
a different set of rectification phases occurs due to the
asymmetry of the L-shaped barriers. 
Reprinted with permission from
C. Reichhardt and C. J. Olson Reichhardt,
Phys. Rev. E {\bf 88}, 042306 (2013). Copyright 2013 by the American Physical
Society.
}
\label{fig:13}
\end{figure}

It is also possible to obtain ratchet effects for circle swimmers moving over asymmetric
substrates \cite{79}.
Figure~\ref{fig:12}(a) shows a counterclockwise swimming particle moving with
a motor force amplitude of $A=0.57$ over an array of L-shaped barriers.
The particle translates one  substrate
lattice constant in the positive $y$ direction
during every swimming cycle.
For $A=0.9$ in Fig.~\ref{fig:12}(b), the particle is confined to a closed
non-translating orbit.
In Fig.~\ref{fig:12}(c), the translating orbit at $ A = 1.5$ includes particle motion
in both the positive $x$ and positive $y$ directions.
Figure~\ref{fig:12}(d) shows that at
$A = 2.05$ the particle translates in the negative $x$-direction.
The average drift velocity of the particle in the $y$ direction, $\langle V_y\rangle$,
and in the $x$ direction, $\langle V_x\rangle$, is plotted as a function of $A$ in
Fig.~\ref{fig:13}.
A series of quantized orbits occur in which the particle
translates in a fixed direction over a finite interval of $A$.
Due to the asymmetry of the L-shaped barriers, the locations of the translating
or ratcheting regimes differ for clockwise and counterclockwise swimming particles;
however, in general, swimmers with different chirality move in different directions,
permitting a mixture of chiral swimmers to be separated by the barrier array.
When thermal effects are included,
the additional
noise expands the range of values of $A$
over which rectification occurs; however, the efficiency of the maximum ratchet effect  
is reduced compared to the nonthermal or deterministic limit. 

\begin{figure}[h]
\includegraphics[width=3in]{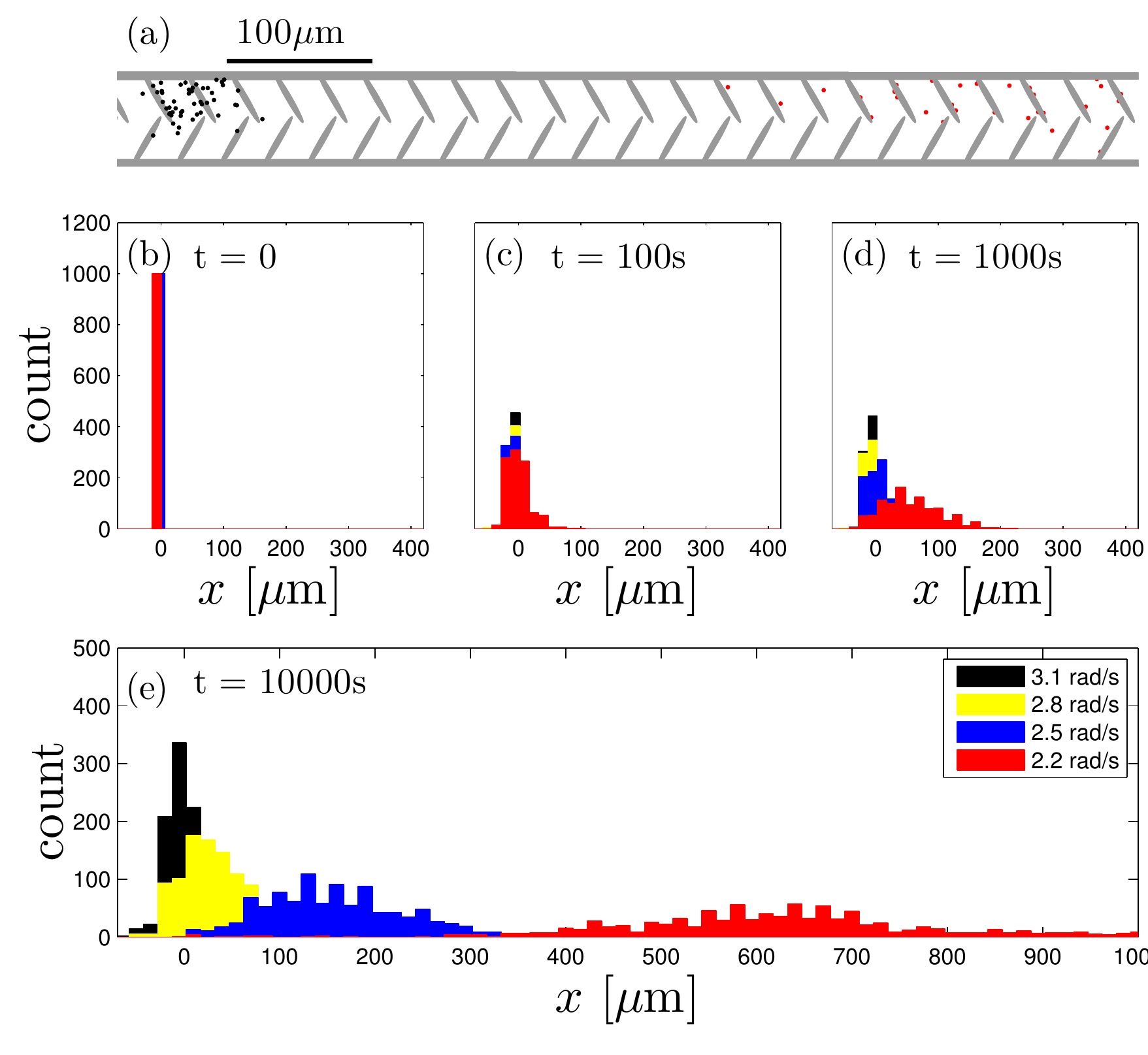}
\caption{ Simulations of
  the separation of circle swimmers with different angular velocities
  by an asymmetric comb channel with the geometry illustrated in panel (a).
  The particles have angular velocities of
  $3.1$ (black), 2.8 (yellow), 2.5 (blue), and $2.2$ (red) rad/sec.
  (b,c,d,e) Histograms of the
  particle positions for increasing times.
  The particles rotating at 2.2 rad/s exhibit the largest amount of rectification.
  Reproduced from
M. Mijalkov and G. Volpe,
Soft Matter {\bf 9}, 6376 (2013), with permission of The Royal Society of Chemistry.
}
\label{fig:14}
\end{figure}

Mijalkov and Volpe \cite{80} considered the separation of
different species of circle swimmers with different angular velocities
by an asymmetric comb channel with a geometry that could be nanofabricated,
as illustrated in Fig.~\ref{fig:14}(a).
The particles are initially placed in the leftmost chamber, and as time
progresses, the particles moving at 2.2 rad/s undergo the largest amount of ratcheting
motion along the channel, as shown by the histograms of particle density at different
times in Fig.~\ref{fig:14}(b-e).
At the longest times of $t = 10000$s
in Fig.~\ref{fig:14}(e) the spatial separation of the different particle species is
clearly visible.

Ai {\it et al.}  \cite{81} considered chiral swimmers interacting with an array 
of half-circle barriers  and found that particles with different chirality migrate to
opposite sides of the sample due to interactions
with the barriers.
In a numerical study of chiral swimmers in a geometry containing M-shaped barriers on the
bottom of the sample at $y=0$ and periodic boundary conditions in the $x$ direction,
the chirality of the swimmers 
leads to a rectification effect and makes it possible to achieve
the transport of a larger passive particle \cite{82}.
The direction in which the active particles ratchet 
depends on the chirality of the swimming, so that
counterclockwise and clockwise swimmers move in opposite directions.
One of the surprising results of this work is
that the direction in which the passive particle drifts can, for some parameters,
be opposite to the direction in which the swimmers are moving,
and this directional reversal
depends strongly on the packing fraction of the system, which again indicates that
collective effects can introduce new  
behaviors.   

\section{VARIANTS ON RATCHET INDUCED TRANSPORT}  
\begin{figure}[h]
  \begin{minipage}{3in}
    \includegraphics[width=3in]{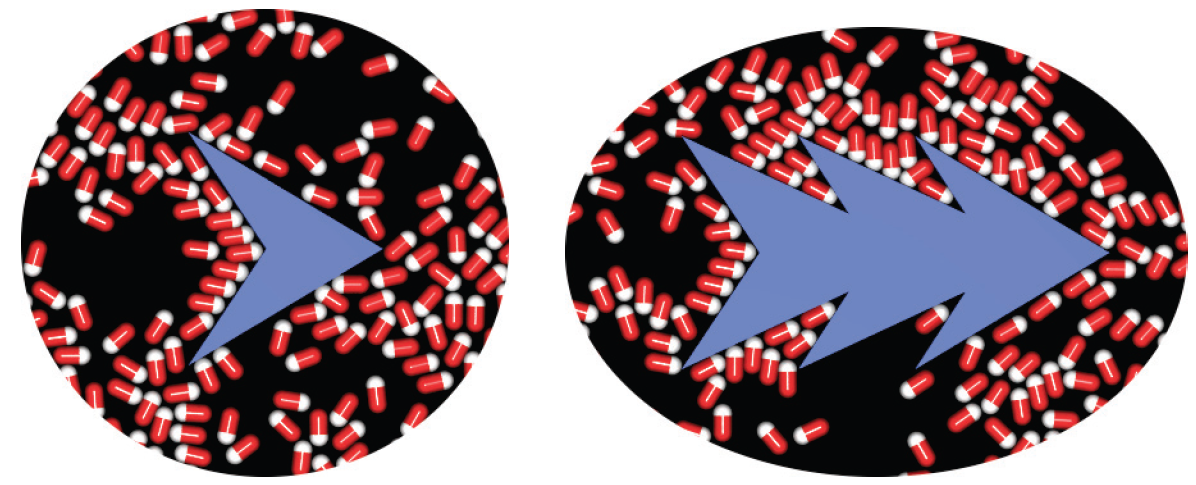}
  \end{minipage}\\
  \begin{minipage}{3in}
    \hspace{0.2in}(a)\hspace{1.0in}(b)
    \end{minipage}\\
  \begin{minipage}{3in}
    \hspace{0.15in}(c)\hspace{-0.05in}
    \includegraphics[width=3in]{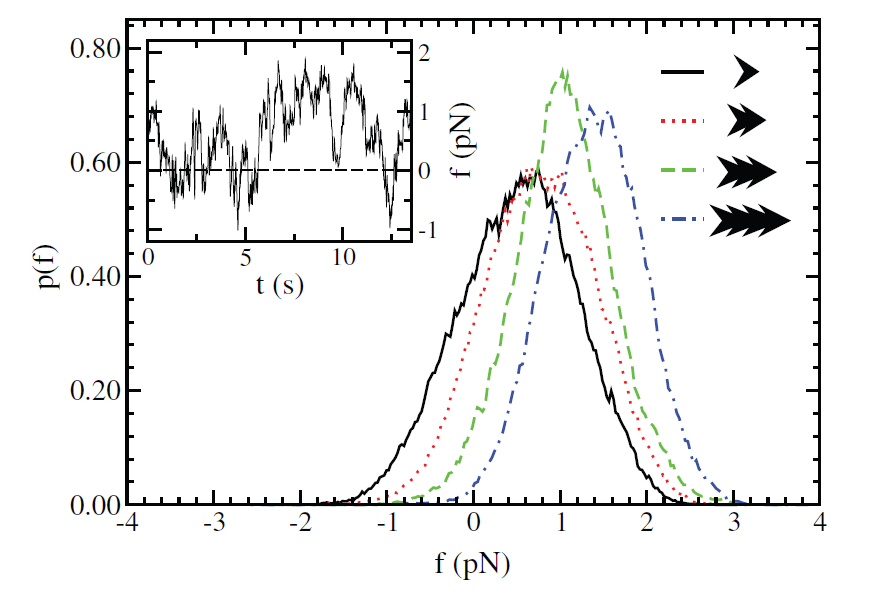}
  \end{minipage}
\caption{
  (a,b) Snapshots from simulations of asymmetric microshuttle objects (blue) interacting
  with self-propelled rods (red/white).  The rods move in the direction of their white tips.
  (a) A shuttle composed of a single wedge shape moves when the active particles
  accumulate along the left side of the wedge.
  (b) Particles accumulate at each corner of a shuttle composed of three wedge shapes.
  (c) The probability distribution $p(f)$ of the time-averaged force
  $f$ exerted on the shuttle by the active particles for shuttles
  made from 1 (black), 2 (red), 3 (green), or 4 (blue) wedges, showing an
  increase in the effective driving force as the number of wedges increases.
  Reprinted with permission from
L. Angelani and R. Di Leonardo,
New J. Phys. {\bf 12}, 113017 (2010).  \textcopyright IOP Publishing \&
Deutsche Physikalishe Gesellschaft.  CC BY-NC-SA.
  http://dx.doi.org/10.1088/1367-2630/12/11/113017
}
\label{fig:15}
\end{figure}

Ratchet effects are generally studied in systems with fixed asymmetric substrates; 
however, if the substrate is mobile and can respond to the active particle,
an active ratchet effect can occur that can be used to 
transport larger-scale  asymmetric objects.
Angelani {\it et al.} \cite{83} examined a
model of  run-and-tumble bacteria interacting with
large scale passive rotary objects that have asymmetric saw-tooth shapes.
The bacteria accumulate in the corners of the sawteeth and generate a net torque
that produces a persistent rotation of the object in one direction.
Such active matter-induced gear rotation
was subsequently confirmed in several experiments with 
swimming bacteria \cite{84,85,86}
as well as for gears interacting with small robots \cite{87}.
Additionally, if an asymmetric object is placed in a
bath of active particles, it is possible to generate enhanced transport
of the object.
Figure~\ref{fig:15}(a,b) shows images 
of asymmetric objects placed in a bath of rod-like
bacteria to create microshuttles \cite{88}. 
The bacteria, which swim in the direction of their white tips, accumulate in the corners
of each shuttle object and produce a net force $f$ that causes the shuttle to translate.
Figure~\ref{fig:15}(c) shows that the probability distribution of $f$ shifts its weight
to higher values of $f$ as the shuttle length is increased, producing a larger
biasing propulsion force.

Experiments on asymmetric wedges in bacterial baths
reveal enhanced transport of these objects \cite{89},
and other studies have examined
how the shape of the asymmetric passive object
affects the transport of the object \cite{90}. 
Since active particles accumulate in corners or concave regions of a funnel
shape, it is possible to use
static \cite{91} or moving wedges \cite{92} to capture 
active particles.      
Many additional variants of using active matter to transport passive objects
are possible.  It would also be interesting
to study systems in which   the passive
objects could change shape in response to the activity. 

\section{CONCLUSION} 
We have highlighted the recent results
for a new class of ratchet system called active ratchets, which appear when the
directed motion of self-propelled particles, which breaks detailed balance, is combined
with an additional asymmetry in the system.
Unlike rocking and flashing ratchets,
active ratchets do not require the application of an external driving force to
produce rectification.
Active ratchet effects and variations upon them will be a
growing field of research
as the ability to fabricate  additional
types of artificial swimmers, nanobots, and other self-driven systems continues
to improve, along with the possibility of applying simple motion rules to the active
objects that can permit them to perform usable work.
Due to the general nature of the conditions under which active ratcheting
can occur,  
it is likely that nature may already be exploiting active ratchet effects in
various biological systems,
so it would be interesting to examine motility process in   
complex biological environments to understand whether active ratchet effects
are present.
Another area for future research is the collective effects that arise in
assemblies of interacting active particles or in more complex objects such as cells.
It has already been demonstrated that collective effects can produce a variety of
novel ratcheting effects including ratchet reversals,
which offers the possibility
to utilize bidirectional ratchets for
various medical technologies and other types of biological applications.

\section*{DISCLOSURE STATEMENT}
The authors are not aware of any affiliations, memberships, funding, or financial holdings that
might be perceived as affecting the objectivity of this review.

\section*{ACKNOWLEDGMENTS}
This work was carried out under the auspices of the 
NNSA of the 
U.S. DoE
at 
LANL
under Contract No.
DE-AC52-06NA25396.

%


\begin{thebibliography}{99}
\bibitem{1}
  Reichhardt C, Reichhardt C J O. 
 arXiv:1602.03798 (unpublished)

\bibitem{2}
  Reimann P. 2002. {\it Phys. Rep.} 361:57--265

\bibitem{3}
  Derenyi I, Vicsek T. 1995. {\it Phys. Rev. Lett.} 75:374--377

\bibitem{4}
  de Souza Silva C C, Van de Vondel J, Morelle M, Moshchalkov V V. 2006.
  {\it Nature (London)} 440:651--654

\bibitem{5}
  Lu Q, Reichhardt C J O, Reichhardt C. 2007.
  {\it Phys. Rev. B} 75:054502

\bibitem{6}
  Lee C S, Jank{\' o} B, Der{\' e}nyi I, Barab{\' a}si A L. 1999.
  {\it Nature (London)} 400:337--340

\bibitem{7}
  Galajda P, Keymer J, Chaikin P, Austin R. 2007.
  {\it J. Bacteriol.} 189:8704--8707

\bibitem{8}
  Reichhardt C, Ray D, Reichhardt C J O. 2015.
  {\it Phys. Rev. B} 91:184502

\bibitem{9}
  Van Oudenaarden A, Boxer S G. 1999.
  {\it Science} 285:1046--1048

\bibitem{10}
  Ertas D. 1998.
  {\it Phys. Rev. Lett.} 80:1548--1551

\bibitem{11}
  Duke T A J, Austin R H. 1998.
  {\it Phys. Rev. Lett.} 80:1552--1555

\bibitem{12}
  Reichhardt C J O, Reichhardt C. 2005.
  {\it Physica C} 432:125--132

\bibitem{12N}
  Dinis L, Perez de Lara D, Gonzalez E M, Anguita J V, Parrondo J M R, Vicent J L. 2009.
  {\it New J. Phys.} 11:073046

\bibitem{14}
  Reichhardt C, Reichhardt C J O. 2016.
  {\it Phys. Rev. B} 93:064508

\bibitem{15}
  Lindner B, Schimansky-Geier L, Reimann P, H{\" a}nggi P, Nagaoka M. 1999.
  {\it Phys. Rev. E} 59:1417--1424

\bibitem{16}
  Reichhardt C, Ray D, Reichhardt C J O. 2015.
  {\it New J. Phys.} 17:070304

\bibitem{17}
  Rousselet J, Salome L, Ajdari A, Prost J. 1994.
  {\it Nature (London)} 370:446--448

\bibitem{18}
  Farkas Z, Tegzes P, Vukics A, Vicsek T. 1999.
  {\it Phys. Rev. E} 60:7022--7031

\bibitem{19}
  Wambaugh J F, Reichhardt C, Olson C J. 2002.
  {\it Phys. Rev. E} 65:031308

\bibitem{20}
  Jones P H, Goonasekera M, Renzoni F. 2004.
  {\it Phys. Rev. Lett.} 93:073904

\bibitem{21}
  Salger T, Kling S, Hecking T, Geckeler C, Morales-Molina L, Weitz M. 2009.
  {\it Science} 326:1241--1243

\bibitem{22}
  Linke H, Humphrey T E, L{\" o}fgren A, Sushkov A O, Newbury R, Taylor R P, Omling P. 1999.
  {\it Science} 286:2314--2317

\bibitem{23}
  Roeling E M, Germs W C, Smalbrugge B, Geluk E J, de Vries T, Janssen R A J, Kemerink M.
  2011.
  {\it Nature Mater.} 10:51--55

\bibitem{24}
  P{\' e}rez-Junquera A, Marconi V I, Kolton A B, {\' A}lvarez-Prado L M, Souche Y, Alija A,
  V{\' e}lez M, Anguita J V, Alameda J M, Mart{\' i}n J I, Parrondo J M R. 2008.
  {\it Phys. Rev. Lett.} 100:037203

\bibitem{25}
  Franken J H, Swagten H J M, Koopmans B. 2012.
 {\it Nature Nanotechnol.} 7:499--503

\bibitem{26}
  Chat{' e} H, Ginelli F, Gr{\' e}goire G, Peruani F, Raynaud F. 2008.
{\it Eur. Phys. J. B} 64:451--456

\bibitem{27}
  Ramaswamy S. 2010.
{\it Annu. Rev. Condens. Matter Phys.} 1:323--345

\bibitem{28}
  Marchetti M C, Joanny J F, Ramaswamy S, Liverpool T B, Prost J, Rao M, Simha R A. 2013.
{\it Rev. Mod. Phys.} 85:1143--1189

\bibitem{29}
  Bechinger C, Di Leonardo R, L{\" o}wen H, Reichhardt C, Volpe G, Volpe G. 
 arXiv:1602.00081 (unpublished).

\bibitem{30}
  Cates M E, Marenduzzo D, Pagonabarraga I, Tailleur J. 2010.
  {\it Proc. Natl. Acad. Sci. (USA)} 107:11715--11720

\bibitem{31}
  Fily Y, Marchetti M C. 2012.
{\it Phys. Rev. Lett.} 108:235702

\bibitem{32}
  Redner G S, Hagan M F, Baskaran A. 2013.
{\it Phys. Rev. Lett.} 110:055701

\bibitem{33}
  Palacci J, Sacanna S, Steinberg A P, Pine D J, Chaikin P M. 2013.
  {\it Science} 339:936--940

\bibitem{34}
  Buttinoni I, Bialk{\' e} J, K{\" u}mmel F, L{\" o}wen H, Bechinger C, Speck T. 2013.
{\it Phys. Rev. Lett.} 110:238301

\bibitem{35}
  Reichhardt C, Reichhardt C J O. 2014.
{\it Soft Matter} 10:7502--7510

\bibitem{36}
  Wan M B, Reichhardt C J O, Nussinov Z, Reichhardt C. 2008.
{\it Phys. Rev. Lett.} 101:018102

\bibitem{37}
  Reichhardt C J O, Drocco J, Mai T, Wan M B, Reichhardt C. 2011.
{\it Proc. SPIE} 8097:80970A

\bibitem{38}
  Tailleur J, Cates M E. 2009.
{\it Europhys. Lett.} 86:60002

\bibitem{39}
  Fily Y, Baskaran A, Hagan M F. 2014.
{\it Soft Matter} 10:5609--5617

\bibitem{40}
  Fily Y, Baskaran A, Hagan M F. 2015.
{\it Phys. Rev. E} 91:012125

\bibitem{48}
  Berdakin I, Jeyaram Y, Moshchalkov V V, Venken L, Dierckx S, Vanderleyden S J,
  Silhanek A V, Condat C A, Marconi V I. 2013.
{\it Phys. Rev. E} 87:052702

\bibitem{41}
  Galajda P, Keymer J, Dalland J, Park S, Kou S, Austin R. 2008.
  {\it J. Mod. Optics} 55:3413--3422

\bibitem{42}
  Hulme E, DiLuzio W R, Shevkoplyas S S, Turner L, Mayer M, Berg H C, Whitesides G M.
  2008.
{\it Lab Chip} 8:1888--95

\bibitem{43}
  Kaehr B, Shear J B. 2009.
{\it Lab Chip} 9:2632--2637

\bibitem{44}
  Kim S Y, Lee E S, Lee H J, Lee S Y, Lee S K, Kim T. 2010.
{\it J. Micromech. Microeng.} 20:085007

\bibitem{45}
  Chen Y-F, Xiao S, Chen H-Y, Sheng Y-J, Tsao H-K. 2015.
{\it Nanoscale} 7:16451--16459

\bibitem{46}
  Angelani L, Costanzo A, Di Leonardo R. 2011.
{\it EPL} 96:68002

\bibitem{New1}
  Yariv E, Schnitzer O. 2014.
{\it Phys. Rev. E} 90:032115

\bibitem{47}
  Potiguar F Q, Farias G A, Ferreira W P. 2014.
{\it Phys. Rev. E} 90:012307

\bibitem{49}
  Berg H C, Brown D A. 1972.
{\it Nature} 239:500--504

\bibitem{50}
  Berdakin I, Silhanek A V, Cort{\' e}z H N M, Marconi V I, Condat C A. 2013.
{\it Central Eur. J. Phys.} 11:1653--1661

\bibitem{51}
  Kantsler V, Dunkel J, Polin M, Goldstein R E. 2013.
{\it Proc. Natl. Acad. Sci. (USA)} 110:1187--1192

\bibitem{52}
  Guidobaldi A, Jeyaram Y, Berdakin I, Moshchalkov V V, Condat C A, Marconi V I,
  Giojalas L, Silhanek A V. 2014.
{\it Phys. Rev. E} 89:032720

\bibitem{53}
  Nam S-W, Qian C, Kim S H, van Noort D, Chiam K-H, Park S. 2013.
{\it Sci. Rep.} 3:3247

\bibitem{N6}
  Ai B-Q, Chen Q-Y, He Y-F, Li F-G, Zhong W-R. 2013.
{\it Phys. Rev. E} 88:062129

\bibitem{N7}
  Wu J C, Chen Q, Wang R, Ai B Q. 2014.
{\it J. Phys. A: Math. Theor.} 47:325001

\bibitem{N8}
  Ai B-Q, Wu J-C. 2014.
{\it J. Chem. Phys.} 140:094103

\bibitem{N9}
  Reichhardt C, Reichhardt C J O. 2013.
{\it Phys. Rev. E} 88:062310

\bibitem{N10}
  Kulic I M, Mani M, Mohrbach H, Thaokar R, Mahadevan L. 2009.
{\it Proc. Roy. Soc. B} 276:2243--2247

\bibitem{N11}
  Pototsky A, Hahn A M, Stark H. 2013.
{\it   Phys. Rev. E} 87:042124

\bibitem{54}
  Stenhammar J, Wittkowski R, Marenduzzo D, Cates M E.
arXiv:1507.01836 (unpublished)

\bibitem{55}
  Nikola N, Solon A P, Kafri Y, Kardar M, Tailleur J, Voituriez R. 
 arXiv:1512.05697 (unpublished)

\bibitem{56}
  Weitz S, Blanco S, Fournier R, Gautrais J, Jost C, Theraulaz G. 2014.
{\it Phys. Rev. E} 89:052715

\bibitem{New4}
  Wu J-C, Chen Q, Wang R, Ai B-Q. 2015.
{\it Physica A} 428:273--278

\bibitem{New5}
  Mijalkov M, McDaniel A, Wehr J, Volpe G. 2016.
{\it Phys. Rev. X} 6:011008

\bibitem{57}
  Drocco J A, Reichhardt C J O, Reichhardt C. 2012.
{\it Phys. Rev. E} 85:056102

\bibitem{58}
  Mahmud G, Campbell C J, Bishop K J M, Komarova Y A, Chaga O, Soh S, Huda S,
  Kandere-Grzyboswka K, Grzybowski B A. 2009.
{\it Nature Phys.} 5:606--612

\bibitem{59}
  Sun X, Driscoll M K, Guven C, Das S, Parent C A, Fourkas J T, Losert W. 2015.
{\it  Proc. Natl. Acad. Sci. (USA)} 112:12557--12562

\bibitem{60}
  Ghosh P K, Misko V R, Marchesoni F, Nori F. 2013.
{\it Phys. Rev. Lett.} 110:268301

\bibitem{61}
  Koumakis N, Lepore A, Maggi C, Di Leonardo R. 2013.
{\it Nature Commun.} 4:2588

\bibitem{62}
  Lambert G, Liao D, Austin R H. 2010.
{\it Phys. Rev. Lett.} 104:168102

\bibitem{63}
  Reichhardt C, Reichhardt C J O. 2014.
{\it Phys. Rev. E} 90: 012701

\bibitem{New3}
  Maass C C, Kr{\" u}ger C, Herminghaus S, Bahr C. 2016
  {\it Annu. Rev. Condens. Matter Phys.} 7:171--193

\bibitem{64}
  Wan M-B, Jho Y-S. 2013.
{\it Soft Matter} 9:3255--3261

\bibitem{65}
  Narayan V, Ramaswamy S, Menon N. 2007.
  {\it Science} 317:105--108

\bibitem{66}
  Sanchez T, Chen D T N, DeCamp S J, Heymann M, Dogic Z.  2012.
 {\it  Nature} 491:431--434

\bibitem{67}
  Giomi L, Bowick M J, Ma X, Marchetti M C.  2013.
{\it   Phys. Rev. Lett.} 110:228101

\bibitem{68}
  DeCamp S J, Redner G S, Baskaran A, Hagan M F, Dogic Z.  2015.
  {\it Nature Mater.} 14:1110--1115

\bibitem{69}
  DiLuzio W R, Turner L, Mayer M, Garstecki P, Weibel D B, Berg H C, Whitesides G M. 2005.
  {\it Nature} 435:1271--1274

\bibitem{70}
  Riedel I H, Kruse K, Howard J. 2005.
  {\it Science} 309:300--303

\bibitem{71}
  Li G, Tam L-K, Tang J X. 2008.
{\it Proc. Natl. Acad. Sci. (USA)} 105:18355--18359

\bibitem{72}
  Tierno P, Johansen T H, Fischer T M. 2007.
{\it Phys. Rev. Lett.} 99:038303

\bibitem{73}
  K{\" u}mmel F, ten Hagen B, Wittkowski R, Buttinoni I, Eichhorn R, Volpe G,
  L{\" owen} H, Bechinger C. 2013.
{\it Phys. Rev. Lett.} 110:198302

\bibitem{74}
  ten Hagen B, K{\" u}mmel F, Wittkowski R, Takagi D, L{\" o}wen H, Bechinger C. 2014.
{\it Nature Commun.} 5:4829

\bibitem{75}
  Reichhardt C, Reichhardt C J O. 2003.
{\it Phys. Rev. E} 68:046102

\bibitem{76}
  Speer D, Eichhorn R, Reimann P. 2009.
{\it Phys. Rev. Lett.} 102:124101

\bibitem{77}
  Tierno P, Johansen T H, Fischer T M. 2007.
{\it Phys. Rev. Lett.} 99:038303

\bibitem{78}
  Nourhani A, Crespi V H, Lammert P E. 2015.
{\it Phys. Rev. Lett.} 115:118101

\bibitem{79}
  Reichhardt C, Reichhardt C J O. 2013.
{\it Phys. Rev. E} 88:042306

\bibitem{80}
  Mijalkov M, Volpe G. 2013.
{\it Soft Matter} 9:6376--6381

\bibitem{81}
  Ai B, He Y, Zhong W. 2015.
{\it Soft Matter} 11:3852--3859

\bibitem{82}
Ai B. 2016.
{\it Sci. Rep.} 6:18740

\bibitem{83}
  Angelani L, Di Leonardo R, Ruocco G. 2009.
{\it Phys. Rev. Lett.} 102:048104

\bibitem{84}
  Di Leonardo R, Angelani L, Dell'Arciprete D, Ruocco G, Iebba V, Schlippa S, Conte M P,
  Mecarini F, De Angelis F, Di Fabrizio E. 2010.
{\it Proc. Natl. Acad. Sci. (USA)} 107:9541--9545

\bibitem{85}
  Sokolov A, Apodaca M M, Grzyboswki B A, Aranson I S. 2010.
{\it Proc. Natl. Acad. Sci. (USA)} 107:969--974

\bibitem{86}
  Kojima M, Miyamoto T, Nakajima M, Homma M, Arai T, Fukuda T. 2016.
{\it Sensors Actuators B} 222:1220--1225

\bibitem{87}
  Li H, Zhang H P. 2013.
{\it EPL} 102:50007

\bibitem{88}
  Angelani L, Di Leonardo R. 2010.
{\it New J. Phys.} 12:113017

\bibitem{89}
  Kaiser A, Peshkov A, Sokolov A, ten Hagen B, L{\" o}wen H, Aranson I S. 2014.
{\it Phys. Rev. Lett.} 112:158101

\bibitem{90}
  Mallory S A, Valeriani C, Cacciuto A. 2014.
{\it Phys. Rev. E} 90:032309

\bibitem{91}
  Kaiser A, Wensink H H, L{\" o}wen H. 2012.
{\it Phys. Rev. Lett.} 108:268307

\bibitem{92}
  Kaiser A, Popowa K, Wensink H H, L{\" o}wen H. 2013.
{\it Phys. Rev. E} 88:022311

\end{thebibliography}
\end{document}